\documentclass{aa}
\usepackage{natbib}

\def\SB9{$S\!_{B^9}$}

\bibpunct{(}{)}{;}{a}{}{,}

\def\vec#1{\mathbf{#1}}

\usepackage[dvips]{graphicx}
\usepackage{times}
\newcommand\ignore[1]{} 
\usepackage{amsmath}

\begin{document}
\title{Proper-motion binaries in the Hipparcos Catalogue}
\subtitle{Comparison with radial velocity data}
\titlerunning{Proper-motion binaries in the Hipparcos Catalogue}
\author{A.~Frankowski
\and S.~Jancart
\and A.~Jorissen\thanks{Senior Research Associate, F.N.R.S., Belgium}}
\institute{Institut d'Astronomie et d'Astrophysique, Universit\'e
Libre de Bruxelles, CP. 226, Boulevard du Triomphe, B-1050
Bruxelles, Belgium   
}

\date{Received date; accepted date}

\abstract
{This paper is the last in a series devoted to the analysis of the
binary content of the Hipparcos Catalogue.}  {The comparison of the
proper motions constructed from positions spanning a short (Hipparcos)
or long time (Tycho-2) makes it possible to uncover binaries with
periods of the order of or somewhat larger than the short time span (in
this case, the 3~yr duration of the Hipparcos mission), since the
unrecognised orbital motion will then add to the
proper motion.}  
{A list of candidate proper motion binaries is
constructed from a carefully designed
$\chi^2$ test evaluating the statistical significance of the difference
between the Tycho-2 and Hipparcos proper motions for 103134 stars in
common between the two catalogues (excluding components of visual
systems).
Since similar lists of proper-motion binaries have already been
constructed, the present paper focuses on the 
evaluation of the detection efficiency   of proper-motion binaries, using
different kinds of control data (mostly radial velocities).
The detection rate
for entries from the Ninth Catalogue of Spectroscopic Binary
Orbits (\SB9) is evaluated, as well as for stars like barium stars,
which are known to be all binaries, and finally for spectroscopic
binaries identified from radial velocity data in the Geneva-Copenhagen
survey of F and G dwarfs in the solar neighbourhood.}  
{Proper motion
binaries are efficiently detected for systems with parallaxes in
excess of $\sim 20$~mas,
and periods in the range 1000 -- 30000~d. The
shortest periods in this range (1000 -- 2000~d, i.e., once to twice the
duration of the Hipparcos mission) may appear only as DMSA/G binaries
(accelerated proper motion in the Hipparcos Double and Multiple System
Annex). Proper motion binaries detected among \SB9\ systems having
periods shorter than about 400~d hint at triple systems, the
proper-motion binary involving a component with a longer orbital
period. A list of 19 candidate triple systems is provided. Binaries
suspected of having low-mass (brown-dwarf-like) companions are listed
as well. Among the 37 barium stars with parallaxes larger than 5~mas,
only 7 exhibit no evidence for duplicity whatsoever (be it
spectroscopic or astrometric). Finally, the fraction of proper-motion
binaries shows no significant variation among the various (regular)
spectral classes, when due account is taken for the detection biases.}
{}
\maketitle

\section{Introduction}
\label{Sect:Intro}

Among the 118\ts218 entries in the Hipparcos Catalogue \citep{ESA-1997},
about 18\ts000 (collected
in the Double and Multiple Systems Annex - DMSA) involve binary or
multiple stars, which break down as follows
\citep[Sect.1.04 of][]{ESA-1997}: 13211 entries are component solutions
(i.e., visual double or multiple systems, DMSA/C), 2622 are acceleration
solutions (i.e., with a variable proper motion, or DMSA/G), 235
astrometric binaries have a known orbit (DMSA/O), 288 are
Variability-Induced Movers (VIMs, or DMSA/V) and finally 1561 are
stochastic solutions (DMSA/X), a number of which might be binaries. With
the exception of  component solutions, all the other kinds of binary
entries have been re-evaluated in the present series of papers
\citep{Pourbaix-03,Platais-03,Pourbaix-2004,Jorissen-2004,Jancart-2005}.
Among the 235 DMSA/O systems, 122 have been re-investigated by
\cite{Jancart-2005} in light of their spectroscopic orbital elements,
and 8 orbital solutions were rejected because of their low quality
\citep[Sect.~4.3 of][]{Jancart-2005}. At the same time, 51 new DMSA/O
were found using their spectroscopic elements published after the
completion of the Hipparcos Catalogue, and collected in the {\it 
Ninth Catalogue of Spectroscopic Binary Orbits}
\citep[\SB9; ][]{Pourbaix-04a}. For the 188 DMSA/V entries that are
long-period variables, the re-processing of their Hipparcos data (using a
variable chromaticity correction) led to the rejection of 161 of them
\citep{Pourbaix-03}. 

Even when spectroscopic orbital elements are not available, for each star from the Hipparcos
Catalogue (with the exception of the DMSA/C component solutions), it is possible to quantify
the likelihood that their {\it Intermediate Astrometric Data} 
\citep[IAD; ][]{vanLeeuwen-1998:a} are better fitted with an orbital model
than with a standard single-star model (i.e., a 5-parameter model,
including the parallax $\varpi$, the $\alpha$ and $\delta$ positions at
the 1991.25 reference epoch, and the proper motion in $\alpha$ and
$\delta$). Hipparcos data are, however, seldom precise enough to derive
the orbital elements from scratch. Therefore, when a spectroscopic orbit is
available beforehand, it is advantageous to import $e, P, T$ from the
spectroscopic orbit and to derive the remaining astrometric elements
\citep[as done by][]{Pourbaix-2000:b,Pourbaix-Boffin:2003,Jancart-2005}.
If a spectroscopic
orbit is not available, trial $(e, P, T)$ triplets scanning a regular
grid (with $10 \le P (\rm d) \le 5000$ imposed by the Hipparcos
scanning law and the mission duration) may be used. In practice, 
it is even possible to avoid
scanning $e$ and $T$, and to restrict the grid to the orbital period
$P$, assuming circular orbits. This choice greatly reduces the
computing time, since test cases revealed that binary systems
with moderate eccentricities are nevertheless detected even
though they are fitted with circular orbits 
\citep[see][ for details]{Pourbaix-2004}. Very eccentric orbits
might be missed, of course, but their small number does not justify
the tremendous increase in computer time required to detect them.
This method has been applied to barium stars by 
\citet{Jorissen-2004,Jorissen-Zacs-2005}, and has been
shown to provide the best detection efficiency if $\varpi > 5$~mas and
$100 \le P({\rm d}) \le 4000$. 

The present paper is the last one in our series reprocessing binaries in
 the Hipparcos Catalogue.
It deals with binaries that are identifiable on the ground of their 
variable proper motion. The DMSA already contains 2622 such  DMSA/G
solutions. This paper aims at
extending   the list of proper-motion binaries\footnote{The
 concept of  proper-motion binaries, which is the topic of the present
 paper,  should not be confused with the more traditional one of  
common proper-motion pairs.} 
by comparing the Hipparcos proper motion with the Tycho-2 proper motion
\citep{Hog-2000:a,Hog-2000:b}, following the idea put forward by
\citet{Wielen-1997} and \citet{Wielen-1998,Wielen-1999}.  
\citet{Kaplan-Makarov-2003}, \citet{Makarov-2004}, and
 \citet{Makarov-Kaplan-2005} presented further developments and
 applications.  The Hipparcos
proper motion, being based on observations  spanning only 3~yrs, may be
altered by the orbital motion, especially for systems with periods in the
range 1500 to 30000~d whose orbital motion was not recognised by
Hipparcos. On the contrary, this effect should average out in the Tycho-2
proper motion, which is derived from observations covering a much longer
time span.  This method
works best when applied to stars with
parallaxes in excess of about 20~mas.
The method is fully described in Sect.~\ref{Sect:Tycho}.

While the present paper was being prepared, \citet{Makarov-Kaplan-2005} 
issued the list of proper-motion binaries at which we were aiming.
\citet{Wielen-1999}  published  a list of proper-motion binaries
(collected in the {\sc DMUBIN} database: 
{http://www.ari.uni-heidelberg.de/datenbanken/dmubin}), 
based on the comparison of Hipparcos proper motions with several other
catalogues, as well.
We nevertheless believe that the present analysis remains timely, for the
following reasons: (i) the statistical method used here to identify
proper-motion binaries is different from the one of
\citet{Makarov-Kaplan-2005}. Our method takes into account the
correlation between the
$\alpha$ and $\delta$ components of the Hipparcos proper motion, as well
as between the Hipparcos and Tycho-2 proper motions, though on a
statistical rather than individual manner for the latter. (ii) We
illustrate the astrophysical promises of the method by applying it to
specific classes of stars like barium stars (Sect.~\ref{Sect:Ba}), Am
stars, or Wolf-Rayet stars (Sect.~\ref{Sect:spectra}). Among those classes,
binaries are sometimes difficult to find using the traditional
spectroscopic approach. For instance, in the case of barium stars,
the periods may be quite long.
(iii) In Sect.~\ref{Sect:SB9}, we test  the detectability
criteria of proper-motion binaries devised by
\citet{Makarov-Kaplan-2005} by looking for such binaries in
the {\it Ninth Catalogue of Spectroscopic Binary Orbits}
\citep[\SB9;][]{Pourbaix-04a}. (iv) The cross-check
between the list of candidate proper-motion binaries and
the corresponding radial-velocity data \citep[available for a large sample
of stars from the HIPPARCOS/CORAVEL survey;][]{Udry-1997} allows us to
identify spurious proper-motion binaries (due to perspective
acceleration) or binaries with low-mass (brown-dwarf-like) companions
(Sect.~\ref{Sect:Gen-Cop}). (v) We evaluate the proportion of
proper-motion binaries among the different spectral classes
(Sect.~\ref{Sect:spectra}) and do not confirm Makarov \& Kaplan's (2005)
suggestion that the fraction of long-period binaries is larger among solar-type 
stars than among the other spectral classes.

%
\section{Statistical method}
\label{Sect:Tycho}
%

The Tycho-2 Catalogue \citep{Hog-2000:a,Hog-2000:b} 
is an extension of the Tycho Catalogue collected by the sky mapper of the ESA 
Hipparcos satellite \citep{ESA-1997}. It contains astrometric positions and
proper motions for $2.5\; 10^6$ stars. 
The proper motions were extracted from the Hipparcos and Tycho-2
catalogues for those stars with only one component, thus excluding the
DMSA/C and DMSA/O entries, as well as Hipparcos objects that have
double entries in the Tycho-2 catalogue.
We ended up with 103\ts134 stars.
 The difference between the Hipparcos
and Tycho-2 proper motions is then evaluated through a $\chi^2$ test:
\begin{equation}
\label{Eq:chi2}
\chi^2= \vec{X}^{\mbox{t}}\; S^{-1}\; \vec{X},
\end{equation}
where $\vec{X}$ is the two-dimensional vector containing the
difference between the Hipparcos and Tycho-2 proper motions, the first
component corresponding to the right ascension $\alpha$ and the second to the 
declination $\delta$. The $2\times2$ covariance matrix $S$ on the proper-motion
differences is obtained from the $4\times4$ covariance matrix $\Sigma$
on the proper motions themselves by the relation
\begin{equation}
S = A \; \Sigma \; A^{\mbox{t}},
\end{equation}
where $A$ is a $2\times4$ matrix such that
\begin{equation}
\vec{X} = A\; \vec{Y} \equiv 
                    \left( \begin{array}{rrrr}
                            1 & 0 & -1 & 0 \\
                            0 & 1 & 0  & -1 \\
                            \end{array}
                    \right) \vec{Y},        
\end{equation}
where $\vec{Y}$ is the 4-dimensional vector containing the $\alpha$
and $\delta$ components  of the Hipparcos proper motion followed by
the $\alpha$ and $\delta$ components of the Tycho-2 proper motion. 
The covariance matrix $\Sigma$ associated with the proper-motion
vector $\vec{Y}$ is expressed by Eq.~\ref{Eq:Sigma}.
\begin{figure*}
\normalsize
\begin{equation}
 \label{Eq:Sigma}
 \Sigma \equiv \left( \begin{array}{cccc}
 \sigma_{\alpha H}^2 & 
 \rho_H   \; \sigma_{\alpha H} \; \sigma_{\delta H} &
 \rho_{HT,\alpha}\; \sigma_{\alpha H} \; \sigma_{\alpha T} &
 0 \\
 \rho_H \; \sigma_{\alpha H} \;  \sigma_{\delta H} &
 \sigma_{\delta H}^2 &
 0 &
 \rho_{HT,\delta} \; \sigma_{\delta H} \; \sigma_{\delta T} \\
 \rho_{HT,\alpha} \; \sigma_{\alpha H} \; \sigma_{\alpha T} &
 0 &
  \sigma_{\alpha T}^2 &
  \rho_T \; \sigma_{\alpha T} \; \sigma_{\delta T} \\
 0 &
 \rho_{HT,\delta} \; \sigma_{\delta H} \; \sigma_{\delta T} &
  \rho_T   \; \sigma_{\alpha T} \; \sigma_{\delta T} &
  \sigma_{\delta T}^2 \\
\end{array} \right).
\end{equation}
\end{figure*}
Here $\sigma_{\alpha H}$ and $\sigma_{\delta H}$ are the standard
errors of the Hipparcos proper-motion $\alpha$ and $\delta$ components,
from fields H17 and H18 of the Hipparcos Catalogue, and $\sigma_{\alpha T}$
and $\sigma_{\delta T}$ are the respective standard errors of the Tycho-2
proper motion, from fields T17 and T18 of the Tycho-2 Catalogue.
The correlation coefficient
$\rho_H$ between the $\alpha$ and $\delta$ component
of the Hipparcos proper motion is taken from field H28 of the
Hipparcos Catalogue. The analogous correlation $\rho_T$ for the Tycho-2
catalogue is supposed to be null, given the different approach used to
construct the Tycho-2 catalogue. The correlation coefficients 
$\rho_{HT,\alpha}$
and $\rho_{HT,\delta}$ between the Hipparcos and Tycho-2 proper
motions were evaluated by \citet{Hog-2000:a}, based on a sample of
supposedly single stars.
To quantify these correlations,
these authors define a parameter $R$
(separately for the $\alpha$ and $\delta$ components; when no
ambiguity is possible, the subscript will therefore be omitted, for
the sake of simplicity) as
\begin{equation}
\label{Eq:R}
R^2 = \frac{\sigma_d^2}{\sigma_{H}^2 + \sigma_{T}^2 +
  \sigma_{c}^2},                    
\end{equation}
and 
\begin{equation}
\label{Eq:sigmad}
\sigma_d^2= \sigma_H^2 + \sigma_T^2 -2 \rho_{HT}\; \sigma_H \;
 \sigma_T + \sigma_c^2,
\end{equation}
 $\sigma_c$ being a  `cosmic error'. The quantity $\sigma_c$
in fact accounts
 for the presence of unrecognised astrometric binaries, which
 artificially increases $\sigma_d$ when comparing proper motions from
 two different catalogues \citep{Wielen-1997,Wielen-1998}. 
This is precisely the effect we
 are after. Therefore, including the cosmic error in the evaluation of
 $R$, which is itself used in the $\chi^2$ test (Eq.~\ref{Eq:chi2}),
 would in fact introduce a bias in the method. 
 Therefore, a new quantity $r$ has been evaluated, closely following
 the definition of $R$ (Eq.~\ref{Eq:R}), but without the cosmic error term. 
 It is listed in Table~\ref{Tab:r} and is derived from the $R$ values
 listed in Table~1 of \citet{Hog-2000:a} using the following expression:
\begin{equation}
\label{Eq:r}
r^2 \equiv \frac{\sigma^2_d}{\sigma^2_H + \sigma^2_T} =
\frac{1}{\frac{1}{R^2} - \frac{\sigma_c^2}{\sigma^2_d}}.
\end{equation}

The parameter $R$ was obtained by \citet{Hog-2000:a} for bins of 1
magnitude from $V_T = 7.0$ to $12.0$ (restricted to subsamples of 
non-binary stars), by computing
$\sigma_d$ and the rms values of the internal errors $\sigma_{H}$ and 
$\sigma_{T}$ of the proper
motions given in the two catalogues, plus an estimate of the cosmic error.
It is important to stress at this point that no $R$ factor is
available for the brightest stars ($V_T < 7$). These bright stars were
not considered in the derivation of $R$ as the vast majority of them, 
having accurate
Hipparcos data, determine the systems of the other catalogues, from
which the Tycho-2 proper motions are derived.  

\begin{table}
\caption{\label{Tab:r}
The quantities $r_\alpha$ and $r_\delta$ as a function of the Tycho
magnitude $V_T$ (see Eq.~\protect\ref{Eq:r})} 
\begin{tabular}{lll}
\hline
\hline
$V_T$ & $r_\alpha$ & $r_\delta$ \\
\hline
$< 7.0 $ & 1. & 1. \\
7.0 - 8 & 0.90 & 0.95 \\
8.0 - 9 & 0.92 & 0.94 \\
9.0 - 10 & 0.89 & 0.93 \\
10.0 - 11 &  0.95 & 0.99 \\
11.0 - 12 &  1.03 & 1.09 \\
$> 12.0$ & 0.98  & 1.13 \\
\hline
\end{tabular}
\end{table} 

The parameter $r$ thus expresses the ratio between the standard
deviation  $\sigma_d$ 
of the difference in Hipparcos and Tycho-2 proper motions   
and the root-mean square of the
 internal errors for these two catalogues. 
In the absence of any correlation, $r$ should
 be unity; in the presence of correlation, $\rho_{HT}$ may be
 expressed
\begin{equation}
\rho_{HT} \; \sigma_{H} \; \sigma_{T} = \frac{1}{2} \; (1 - r^2) \; 
(\sigma^2_{H} + \sigma^2_T). 
\end{equation}   

\begin{figure}
\resizebox{\hsize}{!}{\includegraphics{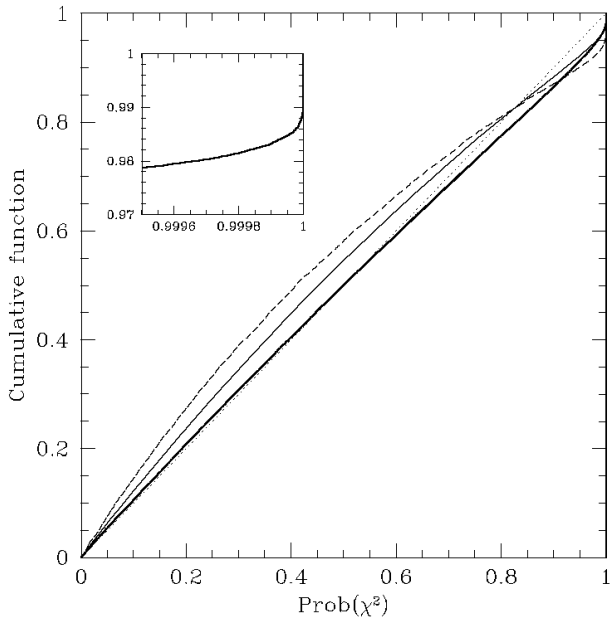}}
\caption{\label{Fig:noR}
Comparison of the cumulative frequency distribution of the $\chi^2$
values from Eq.~\protect\ref{Eq:chi2} evaluated with (thick solid
line) and without (thin solid line) the
correction term $r$ from Eq.~\protect\ref{Eq:r}. Note how well the
$r$-corrected distribution follows the diagonal (dotted line), 
as it should in the absence of binaries. The brightest stars with $V_T
< 7$ were not corrected, and correspond to the dashed line which
deviates markedly from the diagonal. The insert is a zoom on the upper
right portion of the diagram, for the complete sample of 103\ts134 stars.
}
\end{figure}

It should be noted, however, that a value of $r$ different from unity
may also indicate that the `internal errors' $\sigma_H$ or
$\sigma_T$ for the proper motions from the Hipparcos and Tycho-2
catalogues, respectively, were incorrectly evaluated. Equation~\ref{Eq:r}
may indeed be rewritten $\sigma_d^2 = r^2 \sigma_{H}^2 + r^2 \sigma_{T}^2$. 
The two interpretations of $r$ are in fact algebraically
equivalent, as it will become apparent in Eq.~\ref{Eq:S}. 
It is shown in Fig.~\ref{Fig:noR} that the inclusion of $r$ is necessary to
 ensure that the  
 $\chi^2$ variable of Eq.~\ref{Eq:chi2} really follows a $\chi^2$
 distribution. The correctness of our statistical treatment is also
  illustrated in Sect.~\ref{Sect:spectra} in relation with
  Fig.~\ref{Fig:bin_frac}, where the rate of false detections is
  exactly the one expected.

After some algebra, it is then possible to show that  
\begin{equation}
\label{Eq:S}
S \equiv \left( \begin{array}{cc}
  r_\alpha^2 \; (\sigma_{\alpha H}^2 +\sigma_{\alpha T}^2) 
& \rho_H \; r_\alpha \; r_\delta \;\sigma_{\alpha H} \;  \sigma_{\delta H} \\
 \rho_H \; r_\alpha \; r_\delta \; \sigma_{\alpha H} \; \sigma_{\delta H} & r_\delta^2 \; (\sigma_{\delta H}^2 +\sigma_{\delta T}^2 ) 
                         \end{array}\right).
\end{equation}
It must be noted that in the derivation of Eq.~\ref{Eq:S},
to ensure that the determinant of the
$S$ matrix is always positive (ensuring the positiveness of the 
$\chi^2$ as defined in Eq.~\ref{Eq:chi2}), the correlation term 
$\rho_H   \; \sigma_{\alpha H} \; \sigma_{\delta H}$ in
Eq.~\ref{Eq:Sigma} must be replaced by $\rho_H \; r_\alpha \; r_\delta\;
 \sigma_{\alpha H} \; \sigma_{\delta H}$. This modification follows
naturally from the realisation that the $r_\alpha$
and $r_\delta$ factors 
account for the possible mis-evaluation of the internal errors   
$\sigma_{\alpha H}, \sigma_{\delta H}$.

\section{Results: Candidate proper-motion binaries}
\label{Sect:results}

The confidence level of 0.9999 on the $\chi^2$ probability to flag a
star as binary has been 
set to minimise the fraction of false
detections.
This threshold translates into 
$\chi^2 = 18.4206$ (with two degrees of freedom).
The insert in Fig.~\ref{Fig:noR} reveals that 1734 stars
out of 
103\ts134 have Prob$(\chi^2) > 0.9999$\footnote{In the remainder of this paper, 
Prob$(\chi^2)$ denotes the left-sided cumulative $\chi^2$
probability, {\it i.e.}, from 0 to $\chi^2$.} representing 1.68\% of the
total sample, whereas the
expected number of false detections is 0.01\% of 103\ts134, or only 10 stars.
In the remainder
of this paper, stars fulfilling the condition Prob$(\chi^2) > 0.9999$ will
be called (candidate) `$\Delta\mu$ binaries'. Among these 1734 candidate 
binaries, 729 were already flagged in
the Hipparcos DMSA as exhibiting
a non-uniform proper motion (i.e., 162 DMSA/G solutions with 9 parameters: 
the 5 usual astrometric parameters, plus 4 parameters describing the 
first and second derivatives of the proper motion, and 567 DMSA/G
solutions with 7 parameters, corresponding to a linear variation of the proper
motion).

\begin{table*}
\caption{
The first 20 entries in the list of 3565 DMSA/G or $\Delta\mu$ binaries
at the 0.9999 confidence level. The full list is only available
electronically at the CDS, Strasbourg. $\chi^2$ and Prob$(\chi^2)$
characterise the significance of the difference between the Hipparcos
and Tycho-2 proper motions (see Eq.~\protect\ref{Eq:chi2}).  
The `DMSA' column refers to the Double
and Multiple Systems Annex of the Hipparcos Catalogue (ESA 1997):
`5' stands for single-star (5-parameter) solution, `7' and `9' for 
accelerated proper motions (7- and 9-parameter solutions), and  `X' stands 
for stochastic solution.
The columns labelled `Spectrosc. bin.' provide
information about the possible spectroscopic-binary nature of the star,
using data from the
\SB9\ catalogue \citep{Pourbaix-04a}, the Geneva-Copenhagen survey of F
and G dwarfs in the solar neighbourhood
\citep{Nordstrom-2004}, or the KIII-MIII sample of \citet{Famaey-2005}. In
those columns, `B' or `b' means that the star is flagged as a spectroscopic
binary in these catalogues (with Prob$[\chi^2(V_r)] \le 0.001$ and
$0.001 <$ Prob$[\chi^2(V_r)] \le 0.01$, respectively),
`O' that a spectroscopic orbit is available,
`c' that it is present in the catalogue but
that it is not flagged as binary.   The column `Remark' provides a
comment relating to specific properties of the star: long-period variable
(LPV); barium star (Ba);
triple system (triple) or candidate triple system (triple?);
suspected binary with low-mass companion (sB-lm?); spurious
proper-motion binary due to perspective acceleration (perspective acc.). 
\label{Tab:list}
}

\begin{tabular}{rrrccccp{4cm}}
\hline
\hline
HIP & $\chi^2$ & Prob$(\chi^2)$ & DMSA & \multicolumn{3}{c}{Spectrosc. bin.}  & Remark\\
\cline{5-7}
    &          &            &      & {\SB9} & Gen-Cop & K-M \\
\hline
{ 62} &   3.70 & 0.84276 & 7 & - & - & - \\ 
{ 68} & 573.65 & 1.00000 & 7 & - & - & - \\ 
{ 93} &  32.19 & 1.00000 & 5 & - & B & - \\ 
{104} &   9.48 & 0.99126 & 7 & - & - & - \\ 
{171} & 1695.38 & 1.00000 & X & O & B & - \\ 
{290} &  91.51 & 1.00000 & 5 & - & B & - \\ 
{305} &  42.75 & 1.00000 & 7 & - & c & - \\ 
{329} &  37.94 & 1.00000 & 5 & - & - & - \\ 
{340} &   8.28 & 0.98408 & 9 & - & - & - \\ 
{356} &  43.03 & 1.00000 & 5 & - & B & - \\ 
{359} & 105.95 & 1.00000 & 5 & - & - & - \\ 
{493} &   3.74 & 0.84589 & 7 & - & c & - & sB-lm? \\ 
{611} &  11.68 & 0.99708 & 9 & - & - & - \\ 
{626} &  21.38 & 0.99998 & 5 & - & - & - \\ 
{646} &   2.98 & 0.77418 & 7 & - & - & - \\ 
{648} &   5.74 & 0.94337 & 9 & - & - & - \\ 
{695} &   5.10 & 0.92177 & 7 & - & - & - \\ 
{727} &  43.81 & 1.00000 & 5 & - & - & - \\ 
{741} &  13.24 & 0.99867 & 7 & - & - & - \\ 
{765} &  19.10 & 0.99993 & X & - & - & - \\
\hline
\end{tabular}
\end{table*}

Several more proper-motion binaries actually exist in the Hipparcos
Catalogue, since 1831 among the 2560 DMSA/G stars present in our sample
were not flagged as binaries by the $\chi^2$ test. This is in part
a consequence of
the fact that the proper motion provided by the Hipparcos Catalogue for
DMSA/G stars already includes some correction for the orbital motion, so
that the difference between the long-term (i.e. Tycho-2) and short-term
(i.e. Hipparcos) proper motions is not necessarily significant. 
In total, we thus identify 3565 candidate proper-motion binaries (= 1831
DMSA/G stars not reaching the 0.9999 confidence level + 1734 stars above
that level; among the latter, there are 1734 - 729 = 1005 candidate
proper-motion binaries not already flagged as DMSA/G).
These 3565 objects are listed in Table~\ref{Tab:list},
only available in electronic form at the CDS, Strasbourg.
This number is close
-- albeit not identical -- to the 3783 candidate binaries (among which 1161
are not DMSA/G) identified by \citet{Makarov-Kaplan-2005} in their Tables~1 and
2. Before comparing our results with those of \citet{Makarov-Kaplan-2005}, 
it should be mentioned that among DMSA/X (stochastic) solutions, the fraction of candidate proper-motion binaries (45/1403 or 3.2\%) 
is only slightly larger than among the full sample (1.68\%), suggesting
that long-period binaries are not the major cause of stochasticity.

\section{Comparison with Makarov \& Kaplan results}
\label{Sect:Makarov}

\begin{figure}
\resizebox{\hsize}{!}{\includegraphics{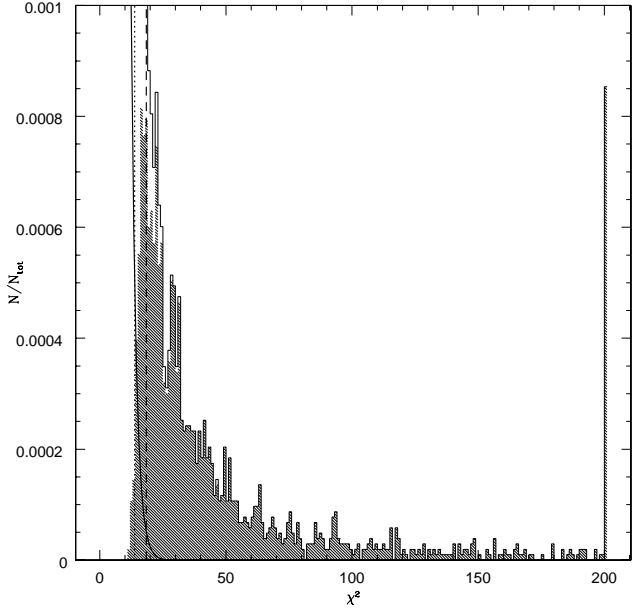}}
\caption[]{\label{Fig:Makarov} Tail of the $\chi^2$ distribution
containing the proper-motion binaries, located to the right of the
vertical dashed line, corresponding to the threshold used
in the present work [Prob$(\chi^2) \ge 0.9999$, translating into $\chi^2
\ge 18.4$ for two degrees of freedom]. 
The candidate proper-motion binaries
from \citet{Makarov-Kaplan-2005} are identified by the shaded
histogram. They are mostly located to the right of the dotted line,
corresponding to Prob$(\chi^2) \ge 0.999$, or $\chi^2 \ge 13.8$ (for two
degrees of freedom). The solid line is the theoretical $\chi^2$
distribution with two degrees of freedom
expected in the case of a pure single-star sample.} 
\end{figure}

\citet{Makarov-Kaplan-2005} adopted a very simple criterion to flag a
star as a candidate proper-motion binary, namely that the difference
between the Hipparcos and Tycho-2 proper motions exceeds 3.5 times
their root-mean square error for at least one of the two
components. For a Gaussian distribution, this threshold corresponds to
a fraction of false detections of $2 \times 4.7\;10^{-4} \sim
10^{-3}$, or 100 stars in a sample of $10^5$. Our threshold is a
factor of 10 more stringent. This is immediately apparent in
Fig.~\ref{Fig:Makarov}, which compares the $\chi^2$ values for Makarov
\& Kaplan's candidate proper-motion binaries (shaded histogram) with
our 0.9999 confidence level corresponding to $\chi^2 \ge 18.4206$.  A 0.999
confidence level
(corresponding to the false detection rate of Makarov \& Kaplan)
translates into $\chi^2 \ge
13.8155$ (for two degrees of freedom), which indeed matches the left
edge of Makarov \& Kaplan's $\chi^2$ distribution of their candidate
proper motion binaries.

It is in the range $12 \le \chi^2 \le 25$ that the two sets differ, as
may in fact be expected (Fig.~\ref{Fig:Makarov}): in that range, 316
stars were flagged as proper-motion binaries by Makarov \& Kaplan, but
not by us, while 138 were flagged by us and not by 
Makarov \& Kaplan. This difference mainly results from the different criteria
used in these two studies: if the difference between the Hipparcos and
Tycho proper motions is about $3.5\sigma$ for one component only, the
corresponding $\chi^2$ will only amount to $3.5^2 = 12.25$, well below
the detection threshold used in the present work. Conversely, the
difference in each component might be slightly smaller than
$3.5\sigma$, but the combined $\chi^2$ (Eqs.~\ref{Eq:chi2} and
\ref{Eq:S}) may be larger than 18.4 (for instance $3.1^2 + 3.1^2 =
19.2$). A good example thereof is {HIP~115929}
not flagged by Makarov \& Kaplan,
despite its $\chi^2$ of 26.12 yielding ${\rm Prob}(\chi^2) =
0.9999978$ in our method. With $\mu_{\alpha*} ({\rm HIP}) =
-5.29\pm1.08$~mas~y$^{-1}$ and $\mu_{\alpha*} ({\rm Tycho-2}) =
-10.2\pm1.2$~mas~y$^{-1}$, the difference in $\mu_{\alpha*}$ amounts
to $|\Delta \mu_{\alpha*}|/ [\epsilon^2({\rm HIP}) + \epsilon^2({\rm
    Tycho-2})]^{1/2} = 3.0$, and with $\mu_{\delta} ({\rm HIP}) = 36.52\pm
0.81$~mas~y$^{-1}$, $\mu_{\delta} ({\rm Tycho-2}) = 32.1\pm
1.1$~mas~y$^{-1}$, the difference in $\delta$ amounts to
$|\Delta\mu_{\delta}|/ [\epsilon^2({\rm HIP}) + \epsilon^2({\rm
Tycho-2})]^{1/2} = 3.2$.
An additional factor acting in the same direction is
that, with a Tycho-2
magnitude $V_T = 9.01$, HIP~115929 belongs to the magnitude range where
the $r_\alpha$ and
$r_\delta$ parameters (Table~\ref{Tab:r}) have the largest impact on
the $\chi^2$ value, increasing it by about 20\%.
Incidentally, HIP~115929 is also flagged as a $\Delta\mu$ binary in the
{\sc DMUBIN} database \citep{Wielen-1999}.

\section{Comparison with other samples of known binaries} 

Our list of candidate proper-motion binaries identified in
Sect.~\ref{Sect:results} can be confronted with several independent
data sets available,
namely
(i) the spectroscopic binaries from the {\it Ninth Catalogue of
Spectroscopic Binary Orbits} \citep[\protect\SB9; ][]{Pourbaix-04a},
(ii) barium stars listed by \citet{Lu-83},
(iii) the radial-velocity standard deviation available from the
Geneva-Copenhagen survey of F and G dwarfs in the solar neighbourhood
\citep{Nordstrom-2004}, and
(iv) the radial-velocity standard deviation available from the
\citet{Famaey-2005} catalogue of K and M giants in the solar neighbourhood.
The latter two data sets are subsamples
resulting from the major effort to collect radial
velocities for about 45\ts000 Hipparcos `survey' stars later than F5
\citep{Udry-1997}, using the CORAVEL spectrovelocimeter
\citep{Baranne-79}.
These data will,  among other things,
enable us to check whether the proper-motion binaries
are indeed found in the orbital-period range predicted by 
\citet{Makarov-Kaplan-2005}
-- an important observational test for the method, not attempted before.
  As a general remark applying to the
  remainder of this paper, we stress that the orbital inclination
  somewhat blurs the comparison between proper-motion and
  spectroscopic diagnostics, since proper-motion binaries seen close to pole-on
  will not be detected as spectroscopic binaries.

\subsection{\protect\SB9\ entries}
\label{Sect:SB9}

\begin{figure}
\resizebox{\hsize}{!}{\includegraphics{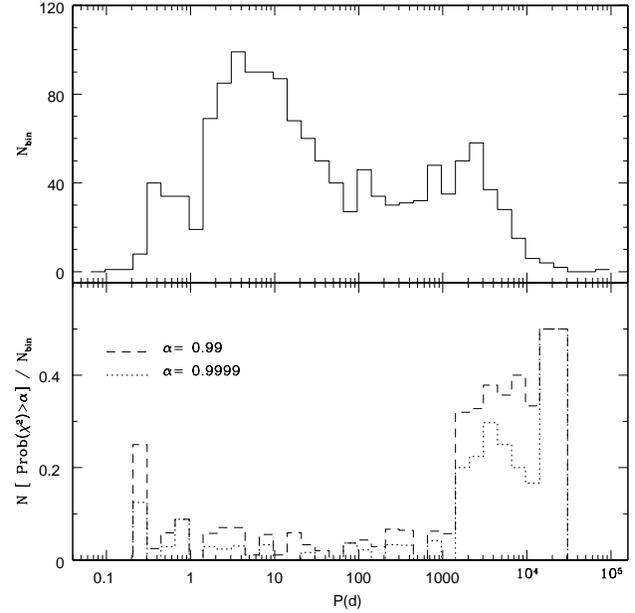}}
\caption[]{\label{Fig:P-SB9}
Upper panel: Orbital-period distribution for stars from the 
\protect\SB9\ catalogue. 
Lower panel: Fraction of stars in
a given bin for which the Hipparcos and Tycho-2 proper motions differ at
confidence levels of 0.9999 (dotted line) and 0.99 (dashed line).}
\end{figure}

\begin{figure}
\resizebox{\hsize}{!}{\includegraphics{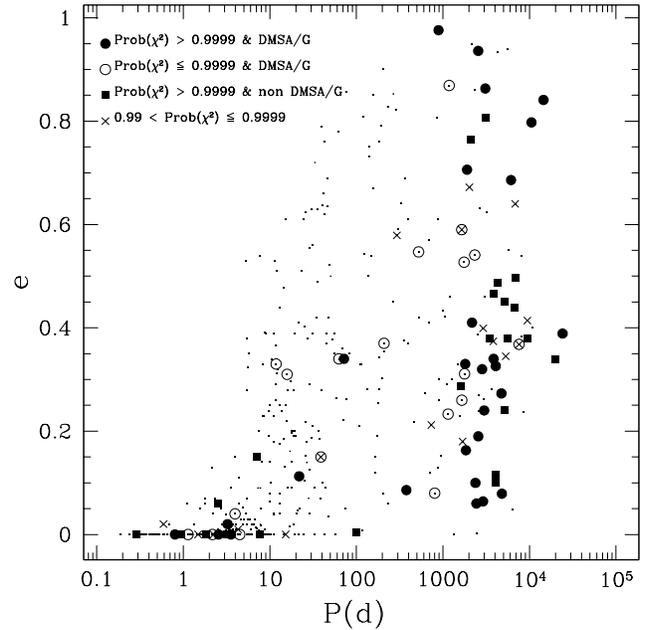}}
\caption[]{\label{Fig:elogP-SB9}
Eccentricity-period diagram for stars from the \protect\SB9\ catalogue
with parallaxes larger than 10~mas.
Symbols are as follows: large filled symbols:
proper-motion binaries detected at the 0.9999 confidence level
(and flagged as DMSA/G: large filled circles;
not flagged as DMSA/G: large filled squares);
open circles: systems flagged as DMSA/G but not fulfilling the 0.9999
confidence level;
crosses: systems not flagged as DMSA/G, lying in the range
0.99 $<$ Prob$(\chi^2) < 0.9999$.
}
\end{figure}

At the time of our query, the {\it Ninth
Catalogue of Spectroscopic Binary Orbits} \citep[\protect\SB9;
][]{Pourbaix-04a} holds orbits for 2386 systems, among which 1320 have a
Hipparcos and Tycho-2 entry. The period distribution for those stars
is presented in Fig.~\ref{Fig:P-SB9}, along with the frequency of
stars for which the Hipparcos and Tycho-2 proper motions differ at
confidence levels of 0.9999 and 0.99.  
Careful inspection of the data shows that the lower limit on period
for efficient $\Delta \mu$ binary detection is located at 1500~d (see also
Fig.~\ref{Fig:elogP-SB9}). Proper-motion binaries with periods up to 30\ts000~d
have been detected, corresponding to the longest periods present in the \SB9\
catalogue.
Among the  long-period proper-motion binaries detected,
all eccentricities are equally well represented.
Figure~\ref{Fig:elogP-SB9} presents an eccentricity-period diagram for
\SB9\ stars with parallaxes larger than 10~mas.
Note that almost all such objects with $P > 1000$~d have been
flagged as either DMSA/G or $\Delta\mu$ binaries.
On the
other hand, $\Delta\mu$ or DMSA/G binaries with orbital periods shorter
than about 400~d are (candidate) triple systems (see the discussion below
and Tables~\ref{Tab:shortPDMSAG} and \ref{Tab:shortPDMSAGnontriple}).

\citet{Makarov-Kaplan-2005} have presented a detailed analysis of the
relationship between the system's orbital period and eccentricity, and
the probability that such a system be either detected as a proper-motion
binary (i.e., with $\Delta \mu$ significantly different from 0), or as
DMSA/G (i.e., with an acceleration term in the proper motion,
$\dot \mu$). They argued that
systems for which the latter effect is stronger
will be mainly found in the period range 3 to
6~yrs (1000 to 2000~d), whereas
systems detectable as $\Delta \mu$ binaries, but not as $\dot{\mu}$ binaries
will be mainly found among
binaries with even longer periods. This is exactly what is seen in
Fig.~\ref{Fig:elogP-SB9}.
Systems
solely detected as DMSA/G and not as proper-motion binaries are mainly 
restricted to the short-period side of the $P > 1000$~d range.

\begin{figure}[htb]
\resizebox{\hsize}{!}{\includegraphics{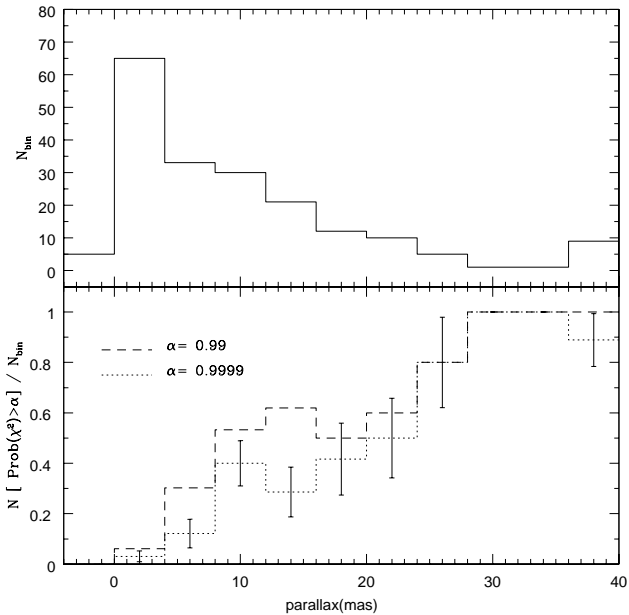}}
\caption[]{\label{Fig:histo-par-SB9}
Upper panel: 
Parallax distribution of \protect\SB9\ stars with orbital periods
larger than 1500~d. Lower panel: Fraction of those \protect\SB9\ stars
detected as proper-motion binaries at confidence levels $\alpha$ of 0.99 and
0.9999. For parallaxes between 10 and 25~mas, about 50\% of the
\protect\SB9\ 
stars with periods larger than 1500~d are detected, and that fraction
becomes larger than 80\% above 25~mas. 
}
\end{figure}

As shown theoretically by \citet{Makarov-Kaplan-2005}, systems ought to 
have a parallax large enough to be detected as proper-motion binaries.
Figure~\ref{Fig:histo-par-SB9} illustrates the incidence of the
  parallax on the detection efficiency: 
for parallaxes between 10 and 25~mas, about 50\% of the
\protect\SB9\ stars with periods larger than 1500~d are detected, and
  that fraction becomes larger than 80\% above 25~mas.
Among the 69 systems
with $P > 1500$~d and parallaxes in excess of 10~mas, 28 
remain undetected,
either as DMSA/G or as proper-motion binaries. 
A closer look at those systems reveals however that 15 of them have
confidence levels in excess of 0.9, and may in fact be considered as
proper-motion binaries.
There is no reason to expect 
that all these 69 binaries should be detected
by the $\Delta \mu$ method, as at 10~mas the detection efficiency of the
method is still below maximum (as shown in Fig.~\ref{Fig:histo-par-SB9} 
and further discussed in Sect.~\ref{Sect:spectra} on a much larger
sample). Nevertheless, we note that among the 13
remaining undetected cases,
 8 might be explained by one of the following reasons:
\begin{itemize}
\item The spectroscopic orbit is of poor quality 
({HIP~108473})\footnote{It must be mentioned that
HIP~108473, not flagged as a $\Delta\mu$ binary when comparing Hipparcos
and Tycho-2 proper motions, is nevertheless flagged as such when comparing
Hipparcos proper motions to the ones computed from the Boss General
Catalogue \citep{Boss-1937,Wielen-1999}.};
\item The system has two visible spectra (SB2 or composite spectrum),
so that the photocentre is not very different from the centre of mass
and no orbital motion can be detected ({HIP~2563},
{HIP~23016}, {HIP~73449},
{HIP~114576}, {HIP~111528});
\item The system is a triple or multiple ({HIP~20430},
HIP~114576);
\item The system is eccentric and was observed by Hipparcos at a
slowly varying orbital phase \citep[{HIP~106786};][]{Gontcharov-2002}.
\end{itemize}   

As opposed to the situation just considered (long-period binaries not
detected as $\Delta\mu$ binaries), systems with $P < 400$~d should not
have been detected as proper-motion binaries.
Unless, that is, the short-period spectroscopic orbit
is in fact the inner orbit of a wider pair involving a third component.
A systematic search for a possible third star in these systems 
has been performed in the {\it Multiple Star Catalogue} 
\citep{Tokovinin-1997} and in the {\it Fourth Catalog of Interferometric
Measurements of Binary Stars} \citep{Hartkopf-2001}, which includes all
published  measurements of binary and multiple star systems obtained by
high-resolution techniques (speckle interferometry, photoelectric
occultation timing, etc.), as well as negative examinations for duplicity.
Table~\ref{Tab:shortPDMSAG} lists the 16 short-period ($P < 400$~d)
DMSA/G
or $\Delta\mu$ systems with firm evidence for being triple or quadruple
hierarchical systems. It is remarkable that about half of the systems
(more precisely 16/39 or 41\%) that are suspected of hosting a third star
from the proper-motion analysis indeed turn out to be triple after close
scrutiny. The case of {HIP~35487} (R~CMa) is exemplary in that respect:
a detailed analysis of a century-long eclipse timing of this system
indeed revealed the presence of a third star \citep{Ribas-2002}. 

Another remarkable case is {HIP~72939}, flagged as a DMSA/G system
but with
a period of 3.55~d listed in \SB9. After the present analysis was completed,
\citet{Fekel-2006} obtained a spectroscopic orbit for a third, outer
companion, with a period of 1641~d. The revised Hipparcos proper motion is
now within 1~$\sigma$ of the Tycho-2 one.

This proportion of 41\% triple systems among short-period \SB9\ binaries
flagged as  DMSA/G or $\Delta\mu$ binaries is much larger than expected
from random selection, since \citet{Halbwachs-2003} find on average 1
triple system for 25-30 systems (3 to 4\%) with  F7-K primaries and
$P < 10$~yr. It therefore seems  reasonable to suggest that the remaining
23
short-period ($P < 400$~d) DMSA/G or $\Delta\mu$ systems will  turn
out to be triple systems as well, despite  the current lack of evidence. They are
listed in Table~\ref{Tab:shortPDMSAGnontriple}. A few stars should perhaps
be removed from this list of candidate triple systems: HIP~21673, HIP~52444,
HIP~68523, and HIP~77911, whose orbital elements  are in fact considered
doubtful (hence, they might not be short-period systems after all).

HIP~77911 deserves further discussion, as it is an outlier in the
eccentricity -- period diagram, with $P = 1.26$~d and $e = 0.61 \pm 0.2$
\citep[][who consider their orbit as very marginal, however]{Levato-1987}.
A visible companion was found by \citet{Kouwenhoven-2005} using the ADONIS
adaptive optics system on the ESO 3.6~m telescope, but it is about 8''
away, and is thus unlikely to be responsible for the DMSA/G nature of the
primary star. Either the orbital period of HIP~77911 is in error, or the
outlying position of HIP~77911 in the eccentricity -- period  diagram
results from eccentricity pumping by a third star in the system
\citep{Mazeh-Shaham-79,Mazeh-90}, the same which is in fact responsible
for the accelerated proper motion.    

Considering the 16 firm and the 19 suspected triple systems, the frequency
of triple systems amount to 35/1320 (or 2.6\%) among the \SB9\ systems
considered in the present analysis. This frequency is consistent with the
frequency of triple systems reported by \citet{Halbwachs-2003}.
To conclude this discussion, it is interesting to remark that several of
the candidate triple stars listed in Table~\ref{Tab:shortPDMSAGnontriple}
are so-called "twins", {\it i.e.}, short-period ($P < 10$~d) SB2 systems with
components of (almost) identical spectral types. It has been argued that
many twins are in fact members of higher-multiplicity systems
\citep[][see however Halbwachs et al. (2003) who do not share this
opinion]{Tokovinin-2000}. \nocite{Halbwachs-2003}

\begin{table*}
\caption[]{\protect\SB9\ 
systems with $P < 400$~d, flagged as DMSA/G or $\Delta\mu$ systems,
and  with firm evidence for being triple or quadruple. 
\label{Tab:shortPDMSAG}}
\begin{tabular}{rcrlp{8cm}}
\hline
\hline
HIP & DMSA/G & $P$ (d) & $e$ & Remark\\
\hline
{24156} & 7 & 0.42 & 0.0 & SB2; triple system \citep{Goecking-1994,Kim-2003}\\  
{30630} & 5 & 6.99 & 0.15 & SB2 resolved; suspected triple as the systemic velocity seems to be
variable\\   
{35487} & 7 & 1.13 & 0.0 & SB2 (F1V + G8IV-V); triple system detected by light-travel time
effect \citep{Ribas-2002} \\
               & 7 & 3.4 $10^4$ & 0.49 \\ 
{42542} & 9 & 48.72 & 0.11  & Triple system \citep{Hartkopf-2001,Pourbaix-04a}\\ 
{49136} & 5 & 0.28 & 0.0  & Quadruple system: two close pairs separated by 0.147 arcsec on the
sky, \\
               & 5 & 0.81 & 0.0 & in an orbit of period $7\;10^3$~d \citep{Tokovinin-1997} \\
{60129} & 7 & 71.9 & 0.34  & Triple system \citep{Hartkopf-1992}\\
               & 7 & 4792 & 0.079 \\
{72939} & 7 & 3.55  & 0.0 & Triple system \citep{Mayor-Mazeh-1987,Tokovinin-1997,Fekel-2006}\\
{84886} & 7 & 21.82 & 0.1126 & Triple system \citep{Griffin-1997,Tokovinin-1997}\\
               & 7 & 2909     &  0.06 \\ 
{86187} & 5 & 2.50 & 0.06 & Triple system: third star found at about 0.3 arcsec;
systemic velocity of the inner pair is variable \citep{Carquillat-1976,Hartkopf-2001}\\ 
{87655} & 7 & 0.80 &  0 & Triple system: the third star is in an orbit of
period 8.9~yrs around the close pair \citep{Soderhjelm-1999,Tokovinin-1997}\\ 
{88848} & X & 1.81 & 0.0   & Triple system \citep{Fekel-2005}\\
               & X & 2092.2 & 0.765 \\
{90312} & 5 & 8.80 & 0 & SB2 (G2IV + K2IV); suspected non single
in Hipparcos\\
{98954} & 9 & 198.7 & 0.05 & A third component at 2.3" \citep{Tokovinin-1997}\\
{103987} & 9 & 377.82 & 0.086 &  Triple system 
\citep[third star at 0.132" with an
estimated period of 14~yrs; ][]{Mason-2001}\\ 
{105091} & 7 & 225.44 & 0.23 & Triple system \\ 
                & 7 & 5.41 & 0.11  \\
{117712} & 5 & 7.75 & 0.0 & Twin SB2: K3V + K3V; triple system:  third star at 4.6" in an orbit of period $9\;10^4$~d \citep{Tokovinin-1997}\\
\hline
\end{tabular}
\end{table*}

\begin{table*}
\caption[]{List of {\it candidate} triple systems. It consists of
\protect\SB9\ systems with $P < 400$~d, flagged as DMSA/G or
$\Delta\mu$ systems, with no firm evidence so far 
of being a triple system.
In column remark, 'not resolved' means that no visual companion is listed
in the {\it Fourth Catalog of Interferometric Measurements 
of Binary Stars} \citep{Hartkopf-2001}.
\label{Tab:shortPDMSAGnontriple}}
\begin{tabular}{llllp{8cm}}
\hline
\hline
HIP & DMSA/G & $P$ (d) & $e$ & Remark\\
\hline 
{3362}  & 7 & 2.17 & 0.0 & SB2 (dM1e + dM1e)\\
{12716} & 5 & 0.95 & 0 & twin SB2: G6 + K0; not resolved\\
{14273} & 7 & 2.73 & 0.0 & SB2\\
{18080} & 5 & 0.51 & 0.07 & W UMa binary\\
{19248} & 7 & 2.56 & 0 & SB2 \\
{23657} & 5 & 1.58 & 0.06 & SB2 (B5V + B5V); unlikely to be $\Delta\mu$ binary because
parallax is very small\\  
{34630} & 7 & 20.63 & 0.296 & SB2\\ 
{45080} & 7 & 6.74 & 0.18 & \\
{50097} & 5 & 3.24 & 0.0 \\
{53295} & 7 & 15.83 & 0.31 & not resolved\\
{70931} & 9 & 11.82 & 0.33 \\
{89601} & 7 & 4.48 & 0.00 & not resolved\\
{95028} & 7 & 208.8 & 0.37 & not resolved\\
{98351} & 7 & 62.88  & 0.34 & not resolved\\
{100142} & 7 & 2.98 & 0.07 &   twin SB2: B2V + B2V;   not resolved\\
{104987} & 5 & 98.82 & 0.0 & SB2 \& visual binary (G2III + A5V)\\
{105406} & 7 & 3.24 &  0.02 & SB2 (F8V + K6V); not resolved\\
{110514} & 7 & 6.72 & 0.178 & twin SB2: F8IV + F7IV\\
{114639} & 7 & 3.96 & 0.04 & SB2: K1IV + F8V; not resolved\\
\medskip\\
\noalign{Weak candidates}\\
{21673}  & 9 & 38.95 & 0.15 & not resolved; orbital elements
considered 'marginal'\\
{52444} & 7 & 51.55 & 0.0 & orbital elements based on small
number of data points \\
{68523} & 9 & 207.36 & 0.55 & not resolved; orbital elements uncertain\\
{77911} & 7 & 1.26  & 0.61 & either triple system as suggested by
eccentricity pumping or doubtful orbital elements (see text)\\
\hline
\end{tabular}
\end{table*}

\subsection{Barium stars}
\label{Sect:Ba}

Barium stars represent a very interesting control sample for the
binary-detection method discussed in this paper, since they are known
to be all binaries with periods in the range 80 -- $10^4$~d
\citep{Jorissen-03}. Therefore, the detection rate of $\Delta\mu$
binaries among barium stars provides a hint on the detection biases.
The catalogue of \citet{Lu-83} contains 163 {\it bona fide} barium
stars with an HIP entry (excluding the supergiants {HD~65699} and
{HD~206778} = $\epsilon$~Peg).
This number drops to 156 after removing DMSA/C and DMSA/O entries.
Only 9 $\Delta\mu$ binaries are detected among barium stars 
at the confidence level 0.9999
 (HIP~13055,
58948, 65535, 71058, 97613, 98431, 104732, 105294, 112821), 5 of which
being moreover flagged as DMSA/G. This rather small number of
detections may appear surprising at first sight, since astrophysical 
considerations require all barium stars to be binaries 
\citep{McClure-80,McClure-83,McClure-Woodsworth-90,Jorissen-03}. 
All but 12 barium stars have parallaxes smaller than 10~mas, though,
which makes the detection of $\Delta\mu$ binaries among barium stars
very inefficient, as
discussed in Sects.~\ref{Sect:SB9} and \ref{Sect:Gen-Cop}.
Among these 12 barium stars with parallaxes larger than
10~mas, 6 are indeed detected as $\Delta\mu$ binaries at the 0.9999
confidence level.

\begin{figure}
\resizebox{\hsize}{!}{\includegraphics{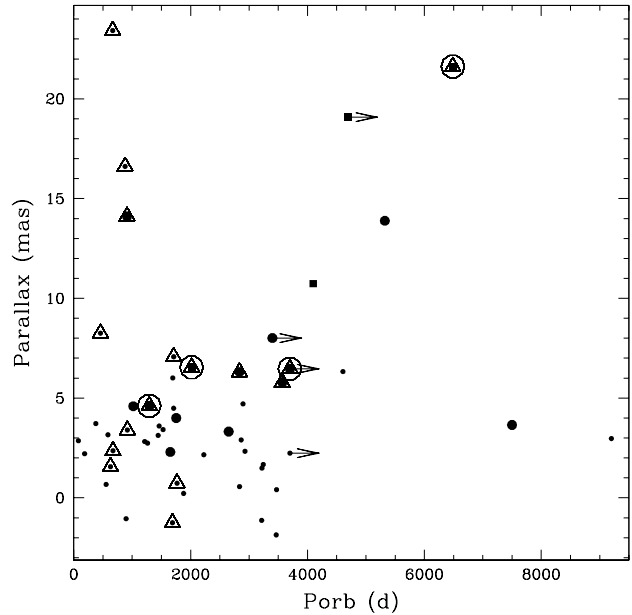}}
\caption{\label{Fig:Ba}
Barium stars with known orbital periods and parallaxes. Objects flagged as
$\Delta \mu$ binaries are represented by filled squares if
Prob$(\chi^2) > 0.9999$ (flag 4 in Table~\ref{Tab:Ba}),
and by large filled circles if $0.9 < {\rm Prob}(\chi^2) \le 0.9999$
(flag 5).
DMSA/G binaries are represented by large open circles. Large open triangles
correspond to binaries where an astrometric orbit could be derived
(flag 3) or at least an astrometric signature of binarity has been
detected (flag 2).
Arrows denote stars with only a lower limit available on the period.
Small filled circles are barium stars that were not flagged as
binaries by any of the astrometric methods to detect duplicity.
}
\end{figure}

\begin{figure}[htb]
\resizebox{\hsize}{!}{\includegraphics{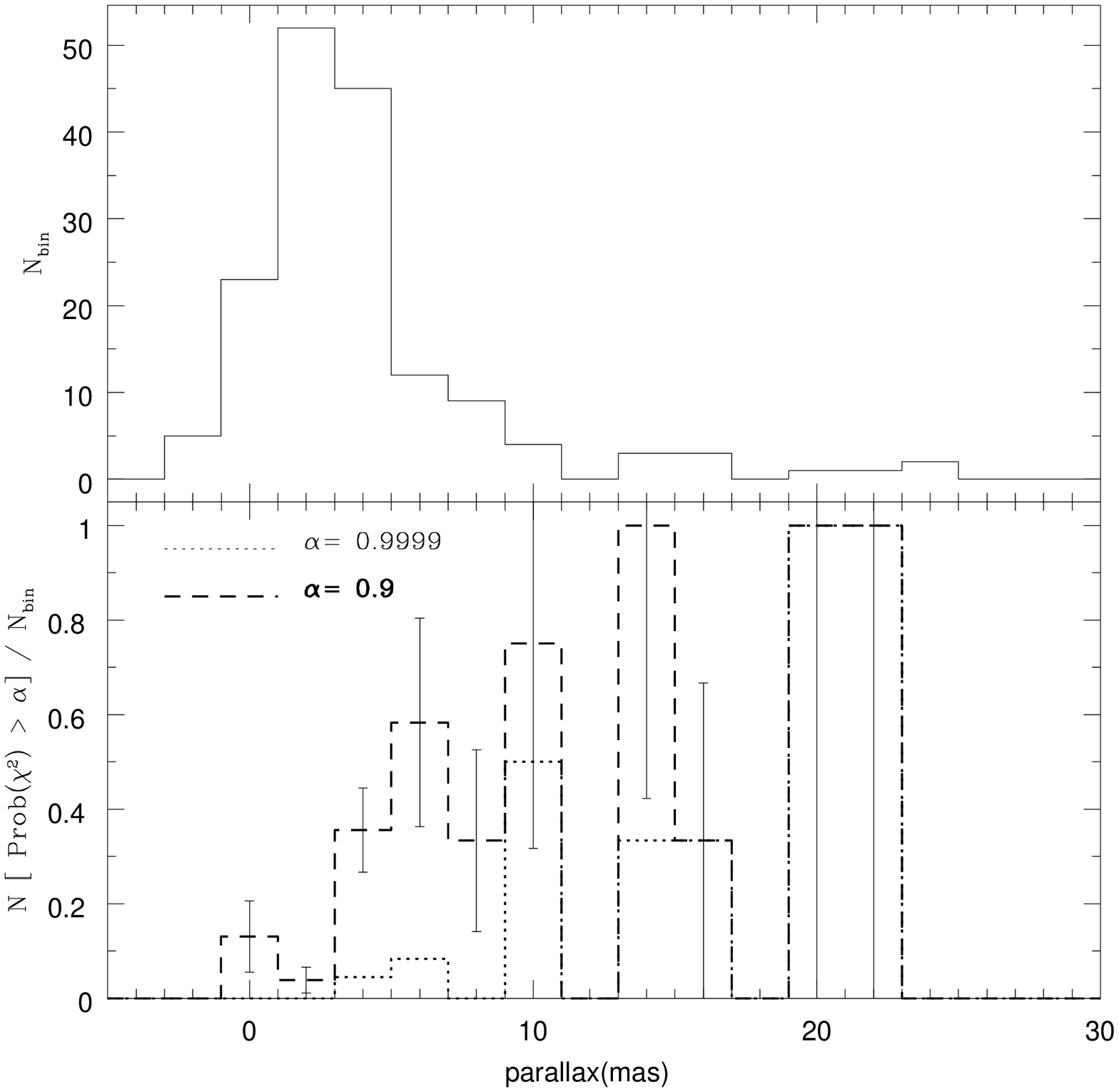}}
\caption[]{\label{Fig:histo-Ba}
Same as Fig.~\protect\ref{Fig:histo-par-SB9}, but for barium stars.
}
\end{figure}

The present astrometric method complements several previous studies
   devoted to the
detection of  binaries among barium stars,  using various astrometric and
spectroscopic methods. Since these studies are
scattered in the literature,
 it is useful to collect their results in one
single table (Table~\ref{Tab:Ba}) and figure (Fig.~\ref{Fig:Ba}). 
Because the efficiency of the
   astrometric methods to detect binaries depends on the parallax, 
Table~\ref{Tab:Ba} has been split in two parts, for barium stars with
parallaxes larger or smaller than 5~mas. This 5~mas
   threshold contrasts with the 20~mas threshold adopted in
   Sect.~\ref{Sect:spectra}. 
A lower threshold for barium stars is motivated by
their scarcity at larger parallaxes (Fig.~\ref{Fig:histo-Ba}).
To reach a detection efficiency of proper-motion binaries of the 
order of 30 to 50\% around 5~mas we relax the confidence level to 0.9
(see the discussion relative to Fig.~\ref{Fig:bin_frac}
in Sect.~\ref{Sect:spectra}; compare with Fig.~\ref{Fig:histo-Ba}).
Objects flagged as $\Delta \mu$ binaries by this less stringent criterion
have a separate flag in Table~\ref{Tab:Ba}.

The flags listed in Table~\ref{Tab:Ba}, referring to various astrometric 
methods, deserve more detailed explanations:
\begin{itemize}
\item Flag 2: The presence of an orbital motion in the astrometric signal
may be checked by
using trial orbital periods and assuming a circular orbit
\citep[for details, see Sect.~\protect\ref{Sect:Intro} and ][]{Pourbaix-2004}, 
even though the spectroscopic orbit is not available.
This method has been applied to barium stars by
\citet{Jorissen-2004,Jorissen-Zacs-2005}, and works best when
$\varpi > 5$~mas and $100 \le P({\rm d}) \le 4000$ (see open triangles in
Fig.~\ref{Fig:Ba});
\item Flag 3: An astrometric orbit constrained by the spectroscopic orbital
elements could be derived
(\citealp{Pourbaix-2000:b}; \citealp{Jancart-2005};
open triangles in Fig.~\ref{Fig:Ba});
\item Flags 4, 5: $\Delta\mu$ binaries, most efficiently detected in
the period range 1500 -- 10000~d.  Flag~4 -- detection with a
confidence level of 0.9999 (filled squares in Fig.~\ref{Fig:Ba}); 
flag~5 -- with a confidence of 0.9 
(large filled circles in Fig.~\ref{Fig:Ba});
\item Flag 6: Binaries detected from their accelerated proper motion
(flagged as DMSA/G in the Hipparcos Catalogue) correspond to periods in the
range {of one} to a few $10^3$~d (large open circles in Fig.~\ref{Fig:Ba}).
\end{itemize}

For the sake of completeness, spectroscopic binaries
\citep[from][]{Jorissen-VE-98} are mentioned as well in Table~\ref{Tab:Ba}
with flag 1 (flag 0 in the case that no radial velocity variations
have been detected by CORAVEL monitoring).
No flag means that the star was not detected as binary by any of the
astrometric methods and it was not included in the CORAVEL survey.
Figure~\ref{Fig:Ba} is remarkable in demonstrating quite clearly the
ability of the various astrometric methods to detect the orbital motion
all the way from 500~d to more than 5000~d, provided that the parallax
be larger than about 5~mas.
The present analysis thus completes the series of papers
\citep{Jorissen-VE-98,Pourbaix-2000:b,Jorissen-2004,Jorissen-Zacs-2005}
devoted to the detection of binaries among barium stars. Among the 37
barium stars with parallaxes larger than 5~mas, there are only 7
with no evidence of orbital motion.

\renewcommand{\baselinestretch}{0.8}
\begin{table}
\caption[]{\label{Tab:Ba}
Duplicity of barium stars with parallaxes $\varpi > 5$~mas. The flags in
the second column correspond to positive (or negative) detections with the
following methods:
(0) no radial-velocity variations detected by CORAVEL monitoring
(1) spectroscopic binary (2) astrometric orbit suspected 
(3) astrometric orbit constrained from spectroscopic orbit (4) 
$\Delta\mu$ binary with Prob$(\chi^2) > 0.9999$ (5) 
$\Delta\mu$ binary with
$0.9 < $~Prob$(\chi^2) \le 0.9999$
(6) accelerated proper motion
(DMSA/G binary). 
Flags (0) and (1) are derived  from
\citet{Jorissen-VE-98}, \citet{Udry-98a}, \citet{Nordstrom-2004}; flag (2) from
\citet{Jorissen-2004,Jorissen-Zacs-2005}; flag (3) from 
\citet{Pourbaix-2000:b}, \citet{Jancart-2005}; flags (4,5) from the present paper;
and flag (6) from the Hipparcos Catalogue DMSA. Parallaxes of barium stars
flagged as $\Delta\mu$ binaries
at a confidence level of 0.9999 (flag 4)
have been written in bold face.}
\begin{tabular}{rlrrl}
\hline
\hline
HIP &Flags & $P_{\rm orb}$  & $\varpi$& Rem. \\ 
    &     &  (d) & (mas) \\
     \hline 
{  944}   &26  & -   &  7.11 \\
{  4422}  &-   & -   & 15.84 \\
{  10119} &25  & -   &  8.91 & Ba dwarf\\ 
{  13055} &12346&2018&  {\bf 6.54} \\ 
{  20102} &1   & 1694&  6.01 \\ 
{  31205} &123 & 457 &  8.25 \\ 
{  32894} &125 & 912 & 14.11 \\  
{  33628} &1256&$>3700$& 6.47 \\ 
{  39426} &05  & -   &  6.34 \\ 
{  47267} &1256& -   &  6.73 & SB in Gen-Cop\\ 
{  49166} &-   & -   &  7.90 \\ 
{  50805} &123 & 667 & 23.42 \\
{  53807} &0   & -   &  9.54 \\ 
{  56130} &-    & -   &  5.46 \\ 
{  56731} &123 & 1711&  7.07 \\ 
{  58948} &14 &$>4700$& {\bf 19.08}\\ 
{  60292} &1235&3569 &   5.78\\ 
{  62608} &5   & -   &  6.20 \\
{  65535} &246 & -   & {\bf 15.73} \\
{  69176} &-   & -   &  8.75 \\
{  71058} &4   & -   &  {\bf 10.63} \\
{  76425} &15  &5324 &  13.89 \\
{  80356} &15&$>3400$&   8.01 \\
{  81729} &-   & -   &  23.35 \\
{  89386} &2   & -   &   5.69 \\
{  89881} &25  & -   &   9.37\\
{  93188} &2   & -   &   5.55 \\
{  98431} &4   & -   &  {\bf 14.06} \\
{ 103263} &1   &4606 &   6.33 \\
{ 104732} &1246&6489 &  {\bf 21.62} \\
{ 105881} &12  &2378 &   8.19 \\
{ 105969} &13  &878  &  16.61 & Ba dwarf \\
{ 106306} &125 &2837 &   6.31 \\
{ 107818} &-   &-    &   7.53 \\
{ 112821} &14  &4098 &  {\bf 10.74} \\
{ 114155} &1   &111  &   6.07 \\
{ 117585} &25  & -   &   7.08 \\
 \hline
 \end{tabular}
 \end{table}
 
 \addtocounter{table}{-1}
 
 \begin{table}
\caption[]{Continued for barium stars with $\varpi < 5$~mas.
Only systems for which at least one flag can be set are listed.
}
\begin{tabular}{rlrrl}
\hline
\hline
HIP &Flags & $P_{\rm orb}$  & $\varpi$& Rem. \\ 
    &     &  (d) & (mas) \\
     \hline
{558  } &5   &   -   &  1.65\\
{21519} &5   &   -   &  4.55\\
{21681} &25  &   -   &  3.44\\
{25452} &15  &  2652 &  3.32\\
{25547} &25  &   -   &  3.43\\
{27767} &5   &   -   &  0.83\\
{29740} &123 &   1689& -1.25\\
{30338} &13  &   629 &  1.56   \\
{32713} &13  &   1768&  0.73\\
{34143} &15  &   7500&  3.65\\
{34795} &02  &   -   &  2.44\\
{36042} &13  &   672 &  2.36\\
{38488} &2   &   -   &  1.29\\
{42999} &5   &   -   &  0.99\\
{47881} &26  &   -   &  3.56\\
{50259} &5   &   -   &  4.38\\
{51533} &5   &   1754&  4.00\\
{52271} &13  &   918 &  3.4\\
{53091} &2   &   -   &  2.14\\
{53717} &15  &  1653 &  2.30\\
{61175} &5   &   -   &  4.89\\
{66844} &05  &   -   &  3.87\\
{88743} &5   &   -   &  3.75\\
{90316} &2   &   -   &  0.56\\
{96024} &2   &   -   &  3.48\\
{97613} &246 &   -   &  {\bf 4.10}\\
{97874} &2   &   -   &  2.07\\
{105294}&246 &   -   &  {\bf 3.49}&  Ba dwarf\\
{107478}&2   &   -   &  3.75\\
{107685}&2   &   -   &  1.77 \\
{109662}&5   &   -   &  -0.10\\
{110108}&15  &  1019 &  4.59\\
{113641}&5   &   -   &  3.17\\    
{117607}&1356&   1294&  4.61\\
{117829}&5   &   -   &  4.42\\
\hline
\end{tabular}
\end{table}

\subsection{Geneva-Copenhagen survey of F and G dwarfs in the
solar neighbourhood}
\label{Sect:Gen-Cop}

The Geneva-Copenhagen catalogue \citep{Nordstrom-2004} 
contains about 16\ts700 F and G dwarfs of the solar neighbourhood.
For 9431 of those, members of visual pairs excluded,
CORAVEL radial velocities as well as Hipparcos and Tycho-2
proper motions are available. It is then possible to confront the
duplicity diagnostic based on proper-motion with that
based on radial velocities. The latter criterion is
based on the comparison between an internal error estimate $\epsilon$
and an external error estimate provided by the standard deviation
$\sigma(V_r)$ of the radial-velocity measurements. From these and the
number of observations $N(V_r)$, the probability Prob$(\chi^2_{V_r})$
that the observed radial-velocity scatter is due to measurement errors
alone may be computed from
$\chi^2_{V_r} = [N(V_r) - 1] \;(\sigma(V_r)/\epsilon)^2$
\citep{Jorissen-Mayor-88}.
We adopt 
$1 - {\rm Prob}(\chi^2_{V_r}) \ge 0.999$
as a threshold for flagging a star with variable
radial velocity, most likely due to duplicity.  
Note that, with the above definitions, 
Prob$(\chi^2_{V_r})$ is the probability for an object of
being {\em single}, while the proper-motion related Prob$(\chi^2)$ 
is the probability for an object of being a binary.

\begin{figure}[]
\resizebox{\hsize}{!}{\includegraphics{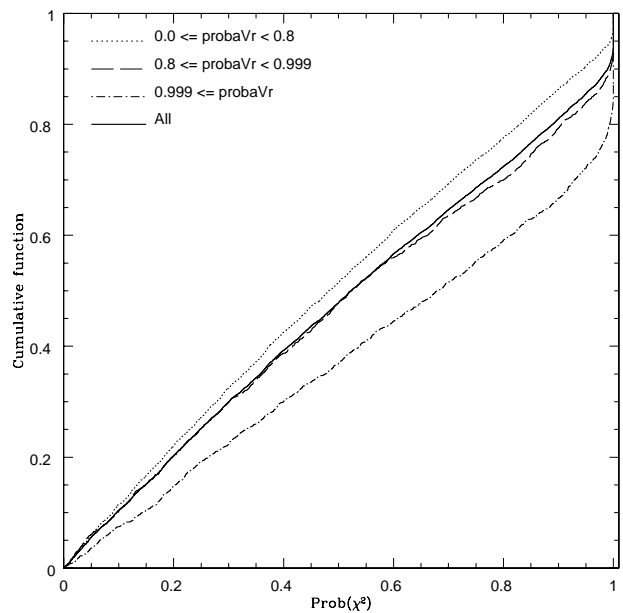}}
\caption[]{\label{Fig:Gen-Cop}
Comparison of the cumulative Prob($\chi^2$) (from Eq.~\protect\ref{Eq:chi2})
for the Geneva-Copenhagen sample of F and G dwarfs, for different values of
their radial-velocity variability probability
(probaVr $\equiv 1 - {\rm Prob}(\chi^2_{V_r})$). The largest
fraction of proper-motion binaries is indeed found among radial-velocity
binaries, having probaVr $\ge 0.999$.
}
\end{figure}

\begin{figure}
\resizebox{\hsize}{!}{\includegraphics{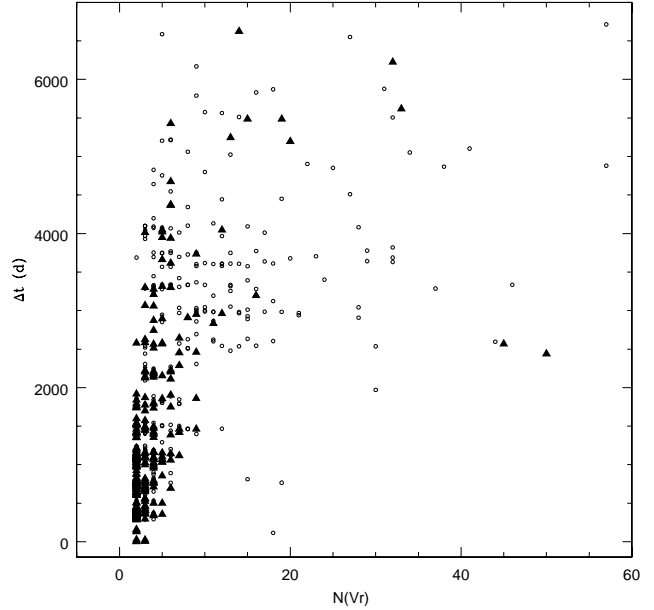}}
\caption[]{\label{Fig:NVrDt} Properties of the radial-velocity data
for stars detected as proper-motion binaries at the 0.9999 level: $N(V_r)$
is the number of radial-velocity data points and
$\Delta t$ their time span (in days).  Open circles correspond to
systems detected as spectroscopic binaries (SB) at the level
Prob$(\chi^2_{V_r}) \le 0.001$, and filled triangles to proper-motion
binaries with constant radial velocity at the level Prob$(\chi^2_{V_r})
> 0.001$.  It is clear that most of the proper-motion binaries not
detected as spectroscopic binaries have either radial-velocity data
points that are scarce ($N(V_r) < 5$), or that  have a short time
span ($\Delta t < 2000$~d).  }
\end{figure}

As revealed by Fig.~\ref{Fig:Gen-Cop}, the largest fraction (about
15\%,
raising to  22\% when adding DMSA/G systems) of proper-motion binaries
indeed arises among the stars flagged as radial-velocity binaries  at the
 99.9\% confidence level. 
These fractions become respectively
   $29.9\pm2.4$\% (= 109/364) and $35.7\pm2.5$\% (= 130/364) 
when eliminating those stars with parallaxes
   smaller than 20~mas among which proper-motion binaries are not
   efficiently detected. 
This fraction of only 35.7\% proper-motion binaries
among spectroscopic binaries reflects the fact that spectroscopic and
proper-motion binaries span different period ranges: 
$P_{\rm PM} \ge 1000$~d for proper-motion binaries
in general, and $P_{\Delta\mu} \ge 1500$~d for $\Delta\mu$
binaries
(see Figs.~\ref{Fig:P-SB9} and \ref{Fig:elogP-SB9}). 
Let $P_{\rm SB}$ be the maximum period of spectroscopic binaries detectable
with the radial-velocity coverage of the Geneva-Copenhagen catalogue,
to be fixed later on.
The relative fractions of those binaries may be computed from the period
distribution of binaries, which has been shown to be log - normal  
\citep{Duquennoy-Mayor-91,Halbwachs-2003,Eggenberger-2004a}: 
\begin{equation}
f(\log P) = \frac{1}{s \sqrt{2 \pi}} e^{-(1/2) z^2},
\end{equation}
where $z = (\log P - 4.8) / s$ and $s = 2.3$, periods being
expressed in days. More precisely, the fraction of binaries with periods in
the range $P_{\rm PM}$ (=~1000~d) -- $P_{\rm SB}$ 
among binaries with periods $P < P_{\rm SB}$ may be
expressed as $1 - {\Phi(z_{\rm PM})}/{\Phi(z_{\rm SB})}$, 
where
\begin{equation}
\Phi(z) = \frac{1}{\sqrt{2\pi}}\int_{-\infty}^z\; {\rm
exp}\left(\frac{-x^2}{2}\right) {\rm d}\,x.
\end{equation}
To obtain the computed fractions
$1 - {\Phi(z_{\rm PM})}/{\Phi(z_{\rm SB})} = 0.36$ and
$1 - {\Phi(z_{\rm \Delta\mu})}/{\Phi(z_{\rm SB})} = 0.29$
consistent with the observed fractions ($0.357\pm0.025$ and
$0.299\pm0.024$, respectively), it is 
required that $z_{\rm SB} = -0.42$ (or $P_{\rm SB} = 6800$~d).
The capability of the
Geneva-Copenhagen survey to detect spectroscopic binaries up to
periods of about 6800~d is not inconsistent
with its typical radial-velocity sampling (see for instance Fig.~\ref{Fig:NVrDt}). 

We now evaluate the total number of proper-motion and spectroscopic binaries
in the Geneva-Copenhagen catalogue.
In total, 1892 stars out of 9431 with HIP and Tycho-2 entries in the
Geneva-Copenhagen catalogue have Prob$(\chi^2_{V_r}) \le 0.001$ and may
thus be considered as {\it bona fide} 
spectroscopic binaries. On top of these, 384
stars are DMSA/G or $\Delta\mu$ binaries (at the 0.9999 level) not
seen as spectroscopic binaries (Prob$(\chi^2_{V_r}) > 0.001$), for a total
of 803 DMSA/G or proper-motion binaries in the Geneva-Copenhagen
catalogue (or 8.5\%).  In total, the fraction of (spectroscopic +
astrometric) binaries in the
Geneva-Copenhagen survey thus amounts to $24.1\pm0.4$\% [= (1892 + 384)/
9431]. 
When restricting the sample to stars with parallaxes in
  excess of 20~mas, this frequency becomes $20.7\pm0.8$\% [= (364 + 131)/
2386].

The 384 proper-motion binaries not detected as spectroscopic
binaries generally have few radial-velocity points ($<5$) not
covering a long time span ($\Delta t < 2000$~d),
as can be seen on Fig.~\ref{Fig:NVrDt}. 
However, thirty systems do not fall in this category:
their radial-velocity data had the required properties to
confirm their binary nature inferred from the proper-motion data,
and yet they are not flagged as spectroscopic binaries at the level 
Prob$(\chi^2_{V_r}) \le 0.001$ in the Geneva-Copenhagen survey. These systems
are listed in Table~\ref{Tab:perspective}, under three different categories: 
(i) single stars spuriously flagged as proper-motion binaries because they exhibit
perspective accelerations, (ii) confirmed spectroscopic or visual binaries, or (iii) stars {\it
suspected} of being  long-period binaries (and seen nearly pole-on, or with low-mass companions, 
or with very eccentric orbits). 
We discuss each category in turn in the remainder of this section. 

The first category corresponds to a purely geometrical effect,
caused by the perspective diminishing (increasing) the proper
motion of a close, fastly receding (respectively approaching)
object. Writing the transverse spatial velocity as $V_t = \mu d
= |V| \cos \beta$ and the radial velocity as $V_r = |V| \sin \beta$  
(where  $d$ is the distance to an object,
$|V| = (V_t^2 + V_r^2)^{1/2}$ is the modulus of the spatial velocity, 
$\beta$ is the angle between the tangential and spatial velocities, such that
$\dot{\beta} = \mu$) and expressing that, for a single star, $|V|$
remains constant over time, yields 
\begin{equation}
\label{Eq:mu}
\dot{\mu} = -2\; \mu\; V_r\; d^{\,-\!1}.
\end{equation}
This perspective acceleration $\dot{\mu}$ may be detected as a secular
change of the proper motion, and confused with a proper-motion binary.
The variation $\Delta \mu$ of the proper motion over the interval
$\Delta t$ may be written as:
\begin{equation}
\label{Eq:perspective}
|\Delta \mu|  =  | 2\; \mu\; V_r\; d^{\,-\!1}\; \Delta t\;
(1 - V_r\; d^{\,-\!1}\; \Delta t)|,
\end{equation}
corresponding to the second-order Taylor expansion of the solution of Eq.~\ref{Eq:mu}.
Perspective acceleration should be detectable
by our $\chi^2$ test with a threshold of 18.4206
for stars having
$ |\Delta \mu| > 4.3\;\sigma_{\Delta \mu}$, where $\sigma_{\Delta \mu}$ is a
typical error of $\alpha$ or $\delta$ component of the proper motion
difference between Hipparcos and Tycho-2. In practice, the situation is
a bit more complicated, since 'instantaneous' proper  motions (as
those appearing in the previous formula) cannot be determined. Proper motions 
are always computed from differences of positions extending over some
time interval, and the effect of perspective acceleration on the positions
is proportional to $\dot{\mu} \; \Delta t^2$. The Tycho-2 proper motion is
derived by a linear least-square fit on these positions spanning
$\Delta t \sim 100$~yr: the Tycho-2 proper motion will therefore fall somewhere in
 the range defined by the 'instantaneous' proper  motions corresponding to each end of
the time interval $\Delta t$, so that
$|\mu_{\rm HIP}  - \mu_{\rm Tycho-2}| < |\Delta \mu|$, with
$|\Delta\mu|$ expressed by Eq.~\ref{Eq:perspective}.

There is no need here, however, to go beyond the simple approximation
expressed by the inequality 
$ | 2\; \mu\; V_r\; d^{\,-\!1}\; \Delta t| > 4.3\;\sigma_{\Delta \mu}$
(with $\Delta t \sim 100$~yr and $\sigma_{\Delta \mu} \sim 1.4$~mas~y$^{-1}$,
as suitable for Hipparcos and Tycho-2 measurements), since
the purpose of the whole discussion is just to identify stars showing
perspective accelerations mistaken for $\Delta\mu$ binaries. These stars
are listed in  Table~\ref{Tab:perspective} \citep[and were already
identified as perspective-acceleration stars in Table~1.2.3 of Volume~1
of the Hipparcos and Tycho Catalogues; ][]{ESA-1997}. For none of the stars listed 
in Table~\ref{Tab:perspective} is there a need to go beyond the second-order Taylor expansion 
considered in Eq.~\ref{Eq:perspective}.

All but four stars in Table~\ref{Tab:perspective} do not, by far, fulfill the criterion for perspective
acceleration. An extensive literature search for these 26 stars resulted in 8 stars with clear
indications for being close visual binaries or spectroscopic binaries. Interestingly enough, two among
these 8 systems have low-mass \citep[HIP~29860;][]{Vogt-2002} or even brown-dwarf companions
\citep[HIP~29295 = Gliese~229;][]{Nakajima-1995}. They constitute further examples of the ability of
the $\Delta\mu$ method to detect binaries with low-mass companions, adding
to the case of {HIP~25647}
already discussed by \citet{Makarov-Kaplan-2005}. In the case of HIP~80902 (=HD~150706), which has been
announced to host a 1 Jupiter-mass exoplanet \citep{Udry-2003}, one may however doubt that the 
$\Delta\mu$ method is able to detect companions that light \citep[see the
discussion of][in relation with their Eq.~2]{Makarov-Kaplan-2005}.
The 18 stars in Table~\ref{Tab:perspective} neither exhibiting
perspective accelerations nor already known as binaries
are flagged candidate binaries with possibly low-mass companions,
although very low inclinations or very eccentric orbits could be other
explanations for their non-detection as spectroscopic binaries by the
Geneva-Copenhagen survey.

\begin{table*}
\caption{\label{Tab:perspective} Stars flagged as DMSA/G or (candidate)
$\Delta\mu$-binaries, not detected as spectroscopic binaries in the
Geneva-Copenhagen survey despite enough radial-velocity data points
($N(V_r) > 5$) spanning more than 2000~d.
For spectroscopic binaries detected elsewhere,
$P_{\rm SB}$ is the spectroscopic period, if applicable.
 If $K \mu$[mas yr$^{-1}$] $|V_r|$ [km
s$^{-1}$] $\varpi$[mas] $> 3.0\;10^{-2}$~mas~yr$^{-2}$,
where $K = 10^{-9}$~[kpc~km$^{-1}$~s~yr$^{-1}$] is a unit conversion factor
to get $\dot\mu$ in mas~yr$^{-2}$, perspective acceleration is detectable.
}
\begin{tabular}{lllllllp{4cm}lll}
\hline
\hline
HIP & Name & $N(V_r)$ & $\Delta t(V_r)$ & Prob$(\chi^2(V_r))$ & $K
\mu_{\rm HIP}\; |V_r|\; \varpi  $ & $P_{\rm SB}$ & Remark \\
    &      &         &   (d)          &       &  (mas~yr$^{-2}$) & (d) \\
\hline
\medskip\\
\noalign{Perspective acceleration}\\
{24186} & Kapteyn & 6 & 4368 & 0.195 & 0.54 & - \\
{54035} & HD 95735 & 19 & 5487 & 0.081 & 0.16 & - & \\
{99461} & HD 191408 & 9 & 2459 & 0.939 & 0.035 & - &\\
{104217}& 61 Cyg B & 32 & 6226 & 0.723 & 0.094 & - & 
\medskip\\
\noalign{Binary systems}\\
{12421} & HD 16619  & 6  & 3940 & 0.280 & $1\;10^{-4}$ & ..  & visual binary ($0.1''$) \\
{29295} & HD 42581 & 15 & 5488 & 0.003 & $5\;10^{-4}$ & .. & Gl 229 B brown dwarf \citep{Nakajima-1995}\\ 
{29860} & HD 43587  & 9 & 2953 & 0.010 & $1\;10^{-4}$ &12309 & $M_{\rm comp} \ge 0.34$~M$_\odot$ \citep{Vogt-2002}\\
{60268} & HD 107582 & 6 & 4676 & 0.805 & $8\;10^{-4}$ & 6864 & SB \citep{Latham-2002}\\
{67275} & $\tau$ Boo & 33 & 5618 & 0.002 & $5\;10^{-4}$ & 3.3 + .. & exoplanet in a wide binary \citep{Patience-2002,Eggenberger-2004b}\\
{80902} & HD 150706  & 12 & 4048 & 0.670  & $8\;10^{-5}$ &264& exoplanetary companion \citep{Udry-2003}\\
{102431}& HD 198084 & 14 & 6625 & 0.036 & $3\;10^{-4}$ & 523 & SB2 \citep{Griffin-1999}\\
{110618}& $\nu$ Ind & 16 & 3196 & 0.612 & $1\;10^{-3}$ & ..  & visual binary ($0.1''$): A3V? + F9
\medskip\\
\noalign{Candidate binaries with low-mass companions (or nearly
  pole-on systems or very eccentric systems)}\\
{493 }  & HD 101   & 7  & 2452 & 0.924  & $2\;10^{-4}$ & - & \\
{4446}  & HD 5357  & 6  & 4375 & 0.128  & $1\;10^{-5}$ & - & \\
{8674}  & HD 11397 & 6  & 2206 & 0.923  & $2\;10^{-4}$ & - & CH subdwarf \citep{Sleivyte-90}\\
{16788} & HD 22309 & 13 & 5247 & 0.045  & $2\;10^{-4}$ & - & non SB in \citet{Latham-2002}\\
{19335} & HD 25998 & 8  & 2908 & 0.234  & $3\;10^{-4}$ & - & no low-mass companion detected with ELODIE \citep{Galland-2005}\\  
{44896} & HD 78612 & 45 & 2567 & 0.008  & $8\;10^{-4}$ & - & \\
{65268} & HD 116316& 7  & 2288 & 0.571  & $8\;10^{-5}$ & - & \\
{66319} & HD 118186 & 50 & 2438 & 0.226 & $8\;10^{-6}$ & - \\
{91215} & HD 171706 & 6  & 2113 & 0.685 & $5\;10^{-5}$ & - \\
{95703} & HD 183473 & 6  & 3615 & 0.003 & $8\;10^{-5}$ & - \\
{98192} & HD 189087 & 20 & 5194 & 0.980 & $3\;10^{-4}$ & - \\
{101726} & HD 196141 & 6 & 3299 & 0.085 & $8\;10^{-5}$ & - \\
{103260} & HD 198829 & 9 & 3734 & 0.015 & $1\;10^{-5}$ & - \\
{103859} & HD 200560 & 12 & 2961 &0.918 & $3\;10^{-4}$ & - \\
{108095} & HD 208068 & 7 & 2644 & 0.932 & $9\;10^{-5}$ & - \\ 
{109461} & LHS 3767  & 11& 2835 & 0.015 & $4\;10^{-4}$ & - \\ 
{111004} & HD 213122 & 6 & 2227 & 0.111 & $2\;10^{-7}$ & - \\ 
{112117} & HD 214953 & 6 & 5429 & 0.714 & $2\;10^{-4}$ & - \\
\hline
\end{tabular}
\end{table*}

\subsection{K and M giants in the solar neighbourhood}
\label{Sect:KM}

\citet{Famaey-2005} published a study of the local kinematics of K
and M giants, accompanied by a catalogue that contains 6691 such
objects, possessing Hipparcos and Tycho-2 proper motions, as well
as CORAVEL radial velocities. The sample can thus be subject to a similar
analysis as done above for the Geneva-Copenhagen catalogue. Again, a
threshold of 0.999 on 1-Prob$(\chi^2_{V_r})$ to identify stars with
variable radial velocities has been adopted. The confrontation of
spectroscopic and $\Delta\mu$ binaries must, however, be restricted to
the 5680 K giants, since in M giants intrinsic radial-velocity variations
(related to envelope pulsations) make the Prob$(\chi^2_{V_r})$
criterion useless to identify spectroscopic binaries \citep[see
Fig.~5 of ][]{Famaey-2005}. 

\begin{figure}[]
\resizebox{\hsize}{!}{\includegraphics{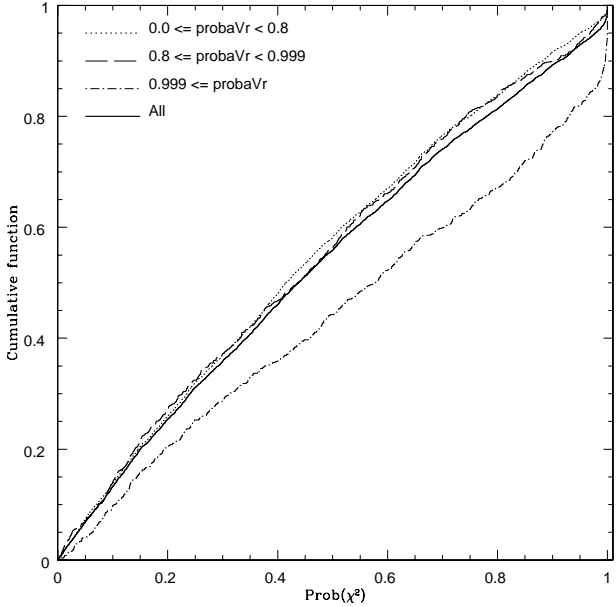}}
\caption[]{\label{Fig:Kgiants}
Same as Fig.~\ref{Fig:Gen-Cop}, but for the K giants from the sample of
\citet{Famaey-2005}. 
The largest
fraction of proper-motion binaries is indeed found among radial-velocity
binaries, having probaVr $\ge 0.999$.
}
\end{figure}

This section presents  an analysis similar to that performed for
the Geneva-Copenhagen sample of F-G dwarfs (Sect.~\ref{Sect:Gen-Cop}; 
compare Fig.~\ref{Fig:Kgiants} 
with Fig.~\ref{Fig:Gen-Cop}) for K giants. The fraction of $\Delta\mu$ binaries
(at the confidence level of 0.9999) among spectroscopic binaries
[Prob$(\chi^2_{V_r}) \le 0.001$] 
amounts to 5.6\% (= 46/816), raising to 21.8\% when adding the
DMSA/G systems. These fractions both become equal to
   $53.8\pm13.8$\% (= 7/13)
when eliminating those stars with parallaxes
   smaller than 20~mas among which proper-motion binaries are not
   efficiently detected. The latter frequency differs by about
   $1.5\sigma$ from the corresponding values 
for the Geneva-Copenhagen sample, and a possible cause for that
difference will be suggested in Sect.~\ref{Sect:spectra}. 
The large errors on the K-giant frequencies
reflect the fact that they are derived on a small sample,
since not many K giants have parallaxes in excess of 20~mas.

We now evaluate the total number of proper-motion and spectroscopic binaries
among K giants. In total, 816 stars out of 5680 have Prob$(\chi^2_{V_r}) \le 0.001$ and may
thus be considered as {\it bona fide} 
spectroscopic binaries. On top of these, 82
stars are DMSA/G or $\Delta\mu$ binaries (at the 0.9999 level) not
seen as spectroscopic binaries (Prob$(\chi^2_{V_r}) > 0.001$), for a total
of 178 DMSA/G or proper-motion binaries.  
The fraction of (spectroscopic +
astrometric) binaries among K giants thus amounts to 15.8\% [= (816 + 82)/
5680]. This fraction raises to $30.2\pm6.3$\% [= (13  + 3)/ 53 ] when
restricting the sample to K giants with parallaxes in excess of
20~mas. This frequency again differs by 
$1.5\sigma$ from the corresponding $20.7\pm0.8$\% frequency
derived for the Geneva-Copenhagen sample.

Among the 82 proper-motion binaries not detected as spectroscopic
binaries, only two ({HIP~24450} and {HIP~78159}) have more
than 5 radial-velocity measurements spanning more than 1000~d, thus with
the potential to detect a possible long-period orbit. Actually, 
HIP~78159 is a visual binary with 1.8'' angular separation and a
period of 41.56~y \citep[about 15\ts000~d; see ][ and references
therein]{Dvorak-1989} which most probably corresponds to the
proper-motion binary. No independent confirmation is available in the
literature about the binary nature of HIP~24450.

\section{Fraction of proper-motion binaries as a function of 
spectral type} 
\label{Sect:spectra}   

\begin{figure}[]
\resizebox{\hsize}{!}{\includegraphics{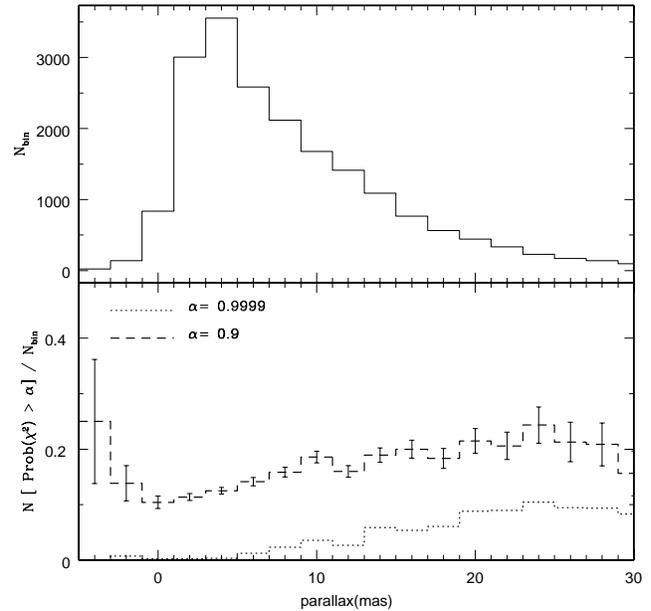}}
\caption[]{\label{Fig:bin_frac}
Upper panel: The parallax distribution of stars with spectral
  type G in the Hipparcos catalogue. Lower panel: Fraction of stars detected as
  $\Delta\mu$ binaries (at confidence levels 0.9 - dashed line - and
  0.9999 - dotted line) in a given parallax bin. Note how the dotted
  curve reaches its final level of 10\% for parallaxes larger than
  20~mas. At 10~mas, only half of the expected systems are detected.
}
\end{figure}

Unbiased estimates of the proportion of binaries among different spectral
types are rare, so that the opportunity offered by the $\Delta\mu$ and
DMSA/G binaries is worth investigating. 
The efficiency of detecting $\Delta\mu$ binaries is affected by the
distance to a tested object. Among farther objects more will be missed by
the method due their smaller (angular) proper motions. To obtain a reliable
estimate of the binary fraction, it is necessary to restrict the sample to
the distance (i.e., parallax) region where the $\Delta\mu$ method attains
its full efficiency. Figure~\ref{Fig:bin_frac} displays the detection fraction as
a function of parallax for G type stars in our complete sample. It can be
seen that a plateau in the detected binary
fraction is reached beyond 20~mas. In the following
analysis of the $\Delta\mu$ binary frequency among the various spectral types, the sample has 
therefore been 
restricted to stars with parallaxes larger than 20~mas. 

\begin{table*} 
\caption{\label{Tab:spectre}
Fraction of proper-motion binaries of a given type among various spectral
classes, normalised by the total number of stars in the considered spectral class 
(labelled $N_{\rm tot}$). Uncertainties on the listed frequencies are
derived from the binomial distribution. The large fluctuations observed in 
Col.~(1) are not physical and 
are caused by the different parallax distributions among the different spectral types 
[see Col.~(6) and text].
} 
\begin{tabular}{lrrllllll} 
\hline 
\hline 
Spectral type & $N_{\rm tot}$ &\multicolumn{1}{c}{(1)} & \multicolumn{1}{c}{(2)} & \multicolumn{1}{c}{(3)} & \multicolumn{1}{c}{(4)} & \multicolumn{1}{c}{(5)} & \multicolumn{1}{c}{(6)} & Catalogue \\ 
\hline
WR-O &251     & 0.012$\pm0.007$  &        $<0.004$  & $< 0.004$        & $<0.004$        & $<0.004$     &  0.4\%\\ 
B   &8664     & 0.022$\pm0.001$  & 0.143$\pm0.132$  & $0.143\pm0.132$  & 0.143$\pm0.132$ & $<0.0001$    &  0.8\%\\ 
A   &15662    & 0.033$\pm0.001$  & 0.152$\pm0.030$  & $0.131\pm0.028$  & 0.083$\pm0.022$ & 0.069$\pm 0.021$ &  0.9\%\\ 
F   &21340    & 0.046$\pm0.001$  & 0.114$\pm0.010$  & $0.091\pm0.009$  & 0.063$\pm0.008$ & 0.051$\pm 0.007$ &  4.3\%\\ 
G   &19628    & 0.043$\pm0.001$  & 0.105$\pm0.008$  & $0.091\pm0.007$  & 0.052$\pm0.006$ & 0.053$\pm 0.006$ &  8.2\%\\ 
K   &29349    & 0.026$\pm0.001$  & 0.093$\pm0.007$  & $0.070\pm0.006$  & 0.051$\pm0.005$ & 0.042$\pm 0.005$ &  6.2\%\\ 
M   &4217     & 0.029$\pm0.002$  & 0.098$\pm0.013$  & $0.077\pm0.012$  & 0.036$\pm0.008$ & 0.061$\pm 0.010$ & 12.3\% 
\medskip\\
All & 103304  & 0.0346$\pm0.0005$& 0.101$\pm0.004$  & $0.080\pm0.004$  & 0.053$\pm0.003$ & 0.048$\pm 0.003$ &  5.3\%
\medskip\\
\hline\\
Am     & 1356 & 0.075$\pm0.007$ & 0.167$\pm0.048$   & $0.133\pm0.044$  & 0.117$\pm0.041$  & 0.050$\pm 0.028$ &  4.4\% & (a)\\
Gen-Cop
(F-G V) &9431 & 0.085$\pm0.003$ & 0.109$\pm0.006$   & $0.089\pm0.006$  & 0.057$\pm0.005$  & 0.052$\pm 0.004$ & 25.3\% & (b)\\
Gliese
(G-K V) &236  & 0.106$\pm0.020$ & 0.109$\pm0.021$   & $0.087\pm0.018$  & 0.043$\pm0.013$  & 0.065$\pm0.016$ &  97.5\% & (c)\\
K III  & 5680 & 0.031$\pm0.002$ & 0.189$\pm0.054$   & $0.189\pm0.054$  & 0.132$\pm0.046$  & 0.057$\pm0.031$ &  0.9\% & (d)\\
Ba &     156  & 0.090$\pm0.023$ & 0.27$\pm0.07^*$   & $0.19\pm0.06^*$  & 0.16$\pm0.06^*$  & 0.11$\pm0.05^*$ &  23.7\%$^*$& (e)\\
   &          &                 & 0.50$\pm0.14^{**}$& $0.50\pm0.14^{**}$& 0.17$\pm0.11^{**}$  & 0.33$\pm0.14^{**}$ &  7.7\%$^{**}$& (e)\\
\hline
\end{tabular} 

Column headings:\\
(1) $\Delta \mu$ (Prob$(\chi^2) > 0.9999$) + DMSA/G (All Prob), all $\varpi$\\ 
(2) $\Delta \mu$ (Prob$(\chi^2) > 0.9999$) + DMSA/G (All Prob), $\varpi > 20$~mas\\ 
(3) $\Delta \mu$ (Prob$(\chi^2) > 0.9999$), $\varpi > 20$~mas\\
(4) DMSA/G, All Prob$(\chi^2)$, $\varpi > 20$~mas\\
(5) Only $\Delta \mu$ (Prob$(\chi^2) > 0.9999$), no DMSA/G (All Prob), $\varpi > 20$~mas\\
(6) Fraction of stars in the considered spectral class with $\varpi > 20$~mas\\
$^*$ For barium stars, these frequencies refer to a parallax threshold of 5 instead of 20~mas\\
$^{**}$ For barium stars, these frequencies refer to a parallax threshold of 10 instead of 20~mas
\medskip\\
Catalogue reference:\\
(a) \citet{Renson-1991}
(b) \citet{Nordstrom-2004}
(c) \citet{Halbwachs-2003}
(d) \citet{Famaey-2005}
(e) \citet{Lu-83}
\end{table*}

Table~\ref{Tab:spectre} lists the fraction of proper-motion binaries
among the various spectral types.
Because $\Delta\mu$ and DMSA/G
binaries probe slightly different period ranges  ($\Delta \mu$
binaries not flagged as DMSA/G, as listed in Col.~5, represent the
long-period tail among proper-motion binaries: see
Fig.~\ref{Fig:elogP-SB9}), they have been listed separately in
Table~\ref{Tab:spectre}. The importance of restricting the sample to
stars with parallaxes larger than 20~mas is very clearly illustrated
by the comparison of Cols.~1 and 2. The fraction of proper-motion
binaries  stays basically constant (within the error bars) among the
various spectral types (Cols.~2 to 5) when restricting the sample to
parallaxes larger than 20~mas, in contrast to the situation prevailing
for the full sample (Col.~1). Since the proportion of stars with
$\varpi > 20$~mas (Col.~6) depends upon the spectral type (because
stars of different spectral types -- and hence average luminosities --
have different average distances in a magnitude-limited sample), the
fraction of the sample in which proper-motion binaries are efficiently
detected varies with spectral type, resulting in the variations
observed in Col.~1 of Table~\ref{Tab:spectre}. Since this selection bias
was not properly recognised by \citet{Makarov-Kaplan-2005}, these authors
concluded that the fraction of
long-period binaries is larger among solar-type stars than among the other
spectral classes (as
is indeed suggested by Col.~1 of Table~\ref{Tab:spectre}).
Based on the above discussion, this conclusion must be discarded, however,
as
shown by the nearly constant frequencies listed in Cols.~2 to 5 of
Table~\ref{Tab:spectre}.
For instance, the fraction of proper-motion binaries among F-G stars is now 
about the same for the Geneva-Copenhagen sample, for the whole Hipparcos
Catalogue and for G and K dwarfs from the Gliese catalogue
\citep{Halbwachs-2003}.

If anything, the fraction of proper-motion binaries seems (slightly) 
larger among spectral types associated with high luminosities, like 
the upper main sequence (B and A types) and the K giants. 
This effect may again result from a selection bias, remembering that
proper-motion binaries cannot be detected in binaries with components
of nearly equal luminosities (since in that case, the photocentre will
be located close to the centre of mass, and there will not be any
detectable orbital motion): in binaries with a high-luminosity primary
star, there is a larger luminosity range accessible for the secondary
(belonging to the lower main sequence), thus increasing the
probability of detecting those systems as proper-motion binaries.  

In the same vein, it should be remarked on Fig.~\ref{Fig:bin_frac} that
there is a large fraction of $\Delta\mu$ binaries among stars with
negative parallaxes (at least when the confidence level is relaxed
from 0.9999 to 0.9), suggesting that the same disturbing effect
(unrecognised orbital motion?) that pushed the astrometric solution
into unphysical territory may have led to (slightly) 
incorrect Hipparcos proper motions. Finally, it is worth nothing that,
for positive parallaxes,
the fraction of $\Delta\mu$ binaries at the 0.9 confidence level runs
almost parallel to the curve corresponding to 0.9999, the two curves
being offset by 10\%, corresponding to the fraction of false
detections expected at the 0.9 confidence level. This constitutes a
further confirmation of the validity of the
statistical treatment described in Sect.~\ref{Sect:Tycho}.

Barium stars are an interesting comparison sample
since they are known to be all binaries (see the discussion in
Sect.~\ref{Sect:Ba}), and yet the frequency of $\Delta\mu$ binaries
listed in Col.~(1) of Table~\ref{Tab:spectre} is only
$0.09\pm0.02$ because of the small proportion of barium stars with large 
parallaxes (only 2 stars 
have parallaxes larger than 20~mas). To get meaningful frequencies, it
has thus been necessary to lower the parallax threshold
adopted for barium stars in Table~\ref{Tab:spectre} 
(see also Sect.~\ref{Sect:Ba}). Two different values have been
adopted: 5~mas, to ensure consistency with the discussion of 
Sect.~\ref{Sect:Ba}, and 10~mas to get a somewhat larger detection
efficiency of proper-motion binaries.     
Although significantly larger -- as expected -- than the frequency of
binaries found in the other spectral classes, this frequency is not
unity because (i) the 5 and 10~mas parallax thresholds only ensure a partial
(about 30 to 50\%; see Figs.~\ref{Fig:histo-par-SB9}, \ref{Fig:histo-Ba},
and \ref{Fig:bin_frac}) detection of the proper-motion binaries; (ii)
as already discussed in Sect.~\ref{Sect:Gen-Cop},
$\Delta\mu$ binaries are restricted to a limited range
in orbital periods (about 1500 -- 30000~d, as clearly apparent from
Fig.~\ref{Fig:elogP-SB9}), whereas barium binaries extend
beyond that period range. 

Am stars are another interesting class to evaluate the frequency of
$\Delta\mu$ binaries, in the framework of a possible relationship
between this class and the barium stars \citep{Hakkila-1989,Jorissen-VanEck-2005}.
The arguments in support of such a connection are based on the binary
nature of stars from both classes, on the similarities of their
abundance peculiarities and of their kinematical properties.  
The main argument to reject any such evolutionary link, however, is based on the significantly
different period ranges observed among Am and barium stars. However,
the question arises whether long-period binaries are really lacking
among Am stars, or whether this lack is a consequence of the
difficulty of identifying them from a radial-velocity monitoring. It is
possible to check this alternative by deriving the fraction of
$\Delta\mu$ binaries among the sample of 1356 Am (and Ap) stars from
the catalogue of \citet{Renson-1991} having an HIP entry.
That fraction is the same as for A-F dwarfs
so that there is no hint for more long-period binaries among Am
stars than among the comparison sample. 

\begin{table*}
\caption[]{\label{Tab:LPVs}
Candidate $\Delta\mu$ binaries at the 0.9999 confidence level or
DMSA/G binaries among long-period variable stars. The Col. labelled
$\varpi$(rev) lists the parallax recomputed as explained in
\citet{Knapp-03}. See text for a word of caution about the reality of these
candidate binaries, inspired by their very small parallaxes.
}
\begin{tabular}{lllrcrcll}
\hline
\hline
HIP & GCVS & Prob($\chi^2$) & $\chi^2$ & DMSA & $\varpi$(HIP) & $\varpi$(rev)  \\
    &     &                &           &     &   (mas)  &      (mas) \\
\hline\\
{10826} & o Ceti & 1.0000000 & 48.66 & 5 & 7.79 & 9.34$\pm$1.07\\
{14042} & T Hor  & 0.9999998 & 30.94 & V & -0.15 & 0.69$\pm$1.22\\
{28874} & S Lep  & 0.9999836 & 22.03 & 5 & 3.63 &  - \\
{34859} & VX Gem & 1.0000000 & 46.72 & 5 & 1.05 & 1.02$\pm$1.99 \\
{78872} & Z Sco  & 0.1404664 & 0.30 &  G & -2.47 & 1.71$\pm$ 2.56 \\
{84948} & RS Her & 0.9309712 & 5.34 &  G & 1.10 & 0.80$\pm$1.41 \\ 
{85617} & TW Oph & 1.0000000 & 37.73 & V & 3.57 & 3.66$\pm$1.17 \\
\hline
\end{tabular}
\end{table*}

\begin{table*}
\caption[]{\label{Tab:WR}
Candidate $\Delta\mu$ binaries at the 0.99 confidence level or DMSA/G
binaries among Wolf-Rayet and O-type stars. The Col. labelled $\varpi$
lists the HIPPARCOS parallax.  SB in Col. 'remark' stands for
'spectroscopic binary'. See text for a word of caution about the reality of
these candidate binaries, inspired by their very small parallaxes.
}
\begin{tabular}{llrrclrl}
\hline
\hline
HIP & HD & Prob$(\chi^2)$ & $\chi^2$ &   DMSA & Sp. type & $\varpi$(HIP) &    Remark\\
    &      &           &     &&&   (mas)  &    \\
\hline\\ 
{48617} & 86161  & 0.9921 & 9.69  & 5 & WR 16 & $-1.01\pm0.74$ & SB$^a$ 10.73 d\\ 
{68995} & 123008 & 0.9963 &11.21  & 5 & O     & $1.95\pm 1.30$\\ 
{89769} & 168206 & 0.9989 &13.62  & 5 & WR 113& $1.38\pm 1.32$ & SB$^b$ 29.1 d  \\ 
{99768} & 192639 & 0.9926 & 9.83  & 5 & O     & $1.22\pm 0.64$\\
{114685}& 219286 & 0.9994 &14.98  & 5 & O7    & $2.04\pm 0.90$\\ 
\hline
\end{tabular}

(a) \citet{Moffat-1982} (b) \citet{Massey-1981,Niemela-1999}
\end{table*}

S stars without lines from the unstable element Tc \citep[also called
{\it extrinsic} S stars;][]{VanEck-98} are a family of binary stars
closely related to barium stars \citep{Jorissen-VE-98,Jorissen-03}. It
is therefore of interest to check the fraction of $\Delta\mu$ binaries
among them as well. The situation is, however, less favourable for
extrinsic S stars than for barium stars because (i) extrinsic S stars
are more luminous (hence they are more distant on average and thus have smaller parallaxes) 
than barium stars \citep{VanEck-98}, and (ii) they are mixed with intrinsic
S stars ({\it i.e.,} genuine asymptotic giant branch stars), which need
not be binaries. Actually, none among the extrinsic S stars in the
Hipparcos Catalogue has a parallax in excess of 10~mas, and it is
therefore not surprising that no $\Delta\mu$ binary is present among
them. The only $\Delta\mu$ binary detected among S stars is {HIP~110487}
($\pi^1$~Gru), an intrinsic S star that is known to be a visual
binary with a G0V companion 2.7'' away \citep{Proust-81}, for which an
orbital period of 6000~yrs is inferred \citep{Knapp-1999}. With such a
long orbital period, one may actually wonder whether the visual
companion is really the one responsible for the $\Delta\mu$ binary, or
whether there could be another, closer companion that plays some role
as well in the shaping of the complex circumstellar environment
inferred from the submm CO lines by \citet{Knapp-1999},
\citet{Chiu-2006}, and \citet{Cruzalebes-2006}.

The search for $\Delta\mu$ binaries among long-period variables (LPVs)
offers an interesting alternative to radial-velocity monitoring since
spectroscopic binaries are very difficult to detect among LPVs,
due to the intrinsic radial-velocity jitter caused by their pulsation
\citep{Jorissen-03}. Table~\ref{Tab:LPVs} reveals a few $\Delta\mu$
and DMSA/G binaries among the 233 LPVs from the Hipparcos Catalogue
collected by \citet{Knapp-03}. These candidate proper-motion binaries are unusual in the
sense that they are detected among stars with very low parallaxes. In
fact, among LPVs with parallaxes in excess of 10~mas (only 3~stars), there are no candidate binaries. This casts doubts on the
reality of the binary detection for stars with parallaxes as small as
those listed in Table~\ref{Tab:LPVs}. Except for o~Cet, there is
unfortunately no independent confirmation of their duplicity.  Parallaxes of very red stars
such as LPVs are subject to specific instrumental effects like
chromaticity, causing a displacement of their stellar image on the
detector (as compared to the image location of a bluer star
occupying the same position on the sky), as discussed by
\citet{Knapp-03} and \citet{Platais-03}. It is possible that this
chromatic effect, if not properly corrected on the older plates
used by the Tycho-2 consortium to derive the proper motion, could
spoil its determination and produce spurious $\Delta\mu$ binaries.

Finally, to conclude this section, it is interesting to list the few
candidate $\Delta\mu$ binaries  found among Wolf-Rayet (WR) and O-type
stars, adding to the spectroscopic and visual binaries listed by
\citet{Mason-1998} and \citet{Niemela-1999}.  
Individual properties of $\Delta\mu$ binaries among WR and O stars, down
to a confidence level of Prob$(\chi^2) > 0.99$, are listed in
Table~\ref{Tab:WR}.
This list may be compared to that of \citet{Hartkopf-1999} and
\citet{Mason-1998}, who searched for visual binaries among O and WR
stars using speckle interferometry. These speckle observations could not
confirm the duplicity of any of the stars of Table~\ref{Tab:WR}. Here as
for LPVs, the small parallaxes cast doubts on the reality of the candidate
binaries.

\section{Conclusions}  
 
The possibility of detecting binaries from the inconsistency between short-span
(like Hipparcos) 
and long-span (like Tycho-2) proper motions \citep[an idea put
forward by][]{Wielen-1997,Wielen-1998,Makarov-Kaplan-2005} 
has been investigated using test samples of known (spectroscopic) binaries.  
Using Hipparcos, Tycho-2, and \SB9\ data, we show that
almost all binaries in the period range 1000 - 30000~d (the lower
bound corresponding to the span of the Hipparcos mission) 
are indeed detected as 'proper-motion binaries'. In the lower range 1000
to 2000~d, binaries are best detected from their accelerated proper
motions (flagged as DMSA/G in the Hipparcos Double and Multiple
Systems Annex). Proper-motion binaries with \SB9\ periods 
below 1000~d are good candidate triple stars, the
proper-motion binary involving a component with a longer orbital
period than the \SB9\ period. A detailed literature search indeed
confirms the triple nature of 16 systems, and a list
of 19 candidate triple systems is provided. 
  
The comparison of short-span and long-span proper motions thus offers a
promising way to detect binaries in extended stellar samples, and has
a bright future given the various upcoming space astrometry missions
like SIM and Gaia. Proper-motion binaries also fill the gap between
spectroscopic and speckle/interferometric binaries. When estimating the
frequency of binaries in a given sample, their contribution is not
negligible: in the Geneva-Copenhagen sample of F and G stars, proper-motion
binaries not detected as spectroscopic binaries add 5.5\% to the fraction
of spectroscopic binaries, 15.2\%, for a total of $20.7\pm0.8$\%. In the
Geneva-Copenhagen sample, there are several proper-motion binaries not
detected as spectroscopic binaries, despite adequate radial-velocity
coverage. These binaries may either be pole-on systems, very eccentric
systems or systems with low-mass companions.

The frequency of proper-motion binaries in fact appears  to be  
the same among all spectral
classes (about $10.1\pm0.4$\% in the whole F-M range),
contrasting with the claim of 
\citet{Makarov-Kaplan-2005} that it is higher among solar-type stars
(because these authors did not account for the selection biases).
There is instead a hint that slightly more proper-motion binaries are
observed for systems with luminous primaries (AV or KIII), as expected if
the light ratio with their companion is generally larger than for systems
hosting primaries on the lower main sequence. Hence, the photocentric motion
is expected to be more frequently detectable -- all other parameters being
equal -- for the former than for the latter.

As expected, barium stars exhibit a significantly larger fraction of
proper-motion binaries (50$\pm15$\% for stars with $\varpi >
10$~mas, consistent with 100\% when duly accounting for the
$\sim50$~\% detection efficiency of proper-motion binaries when
$\varpi\sim 10$~mas). 
There is no hint at a similarly large
fraction of long-period binaries among Am stars; the suspicion of a link
between these two classes must thus be definitely abandoned.   
No convincing candidate proper-motion binaries emerged either among
WR and O stars, or among long-period variable stars.

\begin{acknowledgements} A.F. is a Foreign Postdoctoral Researcher from
FNRS (Belgium) under the grant 1.5.108.05F.
This research was supported in part by an ESA/PRO\-DEX Research Grant 
(15152/01/NL/SFe). 
\end{acknowledgements}

\bibliographystyle{apj} 
\end{document}